\documentclass{aa}  
\usepackage{graphicx}
\usepackage{color}
\usepackage{txfonts}
\newcommand\aproxgt{\mathrel{%
      \rlap{\raise 0.511ex \hbox{$>$}}{\lower 0.511ex \hbox{$\sim$}}}}
\newcommand\aproxlt{\mathrel{%
      \rlap{\raise 0.511ex \hbox{$<$}}{\lower 0.511ex \hbox{$\sim$}}}}

\def\ir1334{{IRAS\,13349+2438}}
\def\mcg6{{MCG--6-30-15}}

\def\apjs{ApJS}
\def\aap{A\&A}

\def\kmps{\ifmmode \rm km~s^{-1} \else $\rm km~s^{-1}$\fi}
\def\psqcm{\ifmmode \rm cm^{-2} \else $\rm cm^{-2}$\fi}
\def\Msun{\ifmmode \rm M_{\odot} \else $\rm M_{\odot}$\fi}
\def\Lsun{\ifmmode \rm L_{\odot} \else $\rm L_{\odot}$\fi}

\newcommand{\qo}{\ifmmode q_{\rm o} \else $q_{\rm o}$\fi}
\newcommand{\Ho}{\ifmmode H_{\rm o} \else $H_{\rm o}$\fi}
\newcommand{\ho}{\ifmmode h_{\rm o} \else $h_{\rm o}$\fi}
\newcommand{\ltsim}{\raisebox{-.5ex}{$\;\stackrel{<}{\sim}\;$}}
\newcommand{\gtsim}{\raisebox{-.5ex}{$\;\stackrel{>}{\sim}\;$}}

\begin{document}

   \title{HST/COS observations of the newly discovered obscuring outflow in NGC 3783}

\author{
G.A. Kriss \inst{1}
\and
M. Mehdipour \inst{2}
\and
J.S. Kaastra \inst{2,3}
\and
A. Rau \inst{4}
\and
J. Bodensteiner \inst{4,5}
\and
R. Plesha \inst{1}
\and
N. Arav \inst{6}
\and
E. Behar \inst{7}
\and 
S. Bianchi \inst{8}
\and
G. Branduardi-Raymont \inst{9}
\and
M. Cappi \inst{10}
\and
E. Costantini \inst{2}
\and
B. De Marco \inst{11}
\and
L. Di Gesu \inst{12}
\and
J. Ebrero \inst{13}
\and
S. Kaspi \inst{7}
\and
J. Mao \inst{2,3}
\and 
R. Middei \inst{8}
\and
T. Miller \inst{6}
\and
S. Paltani \inst{12}
\and
U. Peretz \inst{7}
\and
B.M. Peterson \inst{1,14,15}
\and 
P.-O. Petrucci \inst{16}
\and
G. Ponti \inst{4}
\and
F. Ursini \inst{10}
\and
D.J. Walton \inst{17}
\and
X. Xu \inst{6}
}
\institute{
     Space Telescope Science Institute, 3700 San Martin Drive, Baltimore, MD 21218, USA\\ \email{gak@stsci.edu}
\and SRON Netherlands Institute for Space Research, Sorbonnelaan 2, 3584 CA Utrecht, the Netherlands
\and Leiden Observatory, Leiden University, PO Box 9513, 2300 RA Leiden, the Netherlands
\and Max-Planck-Institut f{\"u}r Extraterrestriche Physik, Gie{\ss}enbachstra{\ss}e, 85748, Garching, Germany
\and Institute of Astronomy, KU Leuven, Celestijnenlaan 200D bus 2401, 3001 Leuven, BE
\and Department of Physics, Virginia Tech, Blacksburg, VA 24061, USA
\and Department of Physics, Technion-Israel Institute of Technology, 32000 Haifa, Israel
\and Dipartimento di Matematica e Fisica, Universit\`{a} degli Studi Roma Tre, via della Vasca Navale 84, 00146 Roma, Italy
\and Mullard Space Science Laboratory, University College London, Holmbury St. Mary, Dorking, Surrey, RH5 6NT, UK
\and INAF-IASF Bologna, Via Gobetti 101, I-40129 Bologna, Italy
\and Nicolaus Copernicus Astronomical Center, Polish Academy of Sciences, Bartycka 18, PL-00-716 Warsaw, Poland
\and Department of Astronomy, University of Geneva, 16 Ch. d'Ecogia, 1290 Versoix, Switzerland
\and European Space Astronomy Centre, P.O. Box 78, E-28691 Villanueva de la Ca\~{n}ada, Madrid, Spain
\and Department of Astronomy, The Ohio State University, 140 West 18th Ave., Columbus, OH 43210, USA
\and Center for Cosmology \& AstroParticle Physics, The Ohio State University, 191 West Woodruff Ave., Columbus, OH 43210, USA
\and Univ. Grenoble Alpes, CNRES, IPAG, 38000 Grenoble, France
\and Institute of Astronomy, Madingley Road, CB3 0HA Cambridge, UK
}

\vspace{1cm}

   \date{Accepted, October 28, 2018}

\vspace{-12pt}
 
  \abstract
  {
  }
   {
   To understand the nature of transient obscuring outflows in active
   galactic nuclei, we use simultaneous multiwavelength observations with
   {\it XMM-Newton, NuSTAR}, the {\it Hubble Space Telescope} (HST),
   and the Max Planck Gesellschaft/European Southern Observatory (ESO)
   2.2-m telescope triggered by soft X-ray absorption detected
   by {\it Swift}.
   }
   {We obtained ultraviolet spectra on 2016 December 12 and 21
   using the {\it Cosmic Origins Spectrograph}
   (COS) on HST simultaneously with X-ray spectra obtained with
   {\it XMM-Newton} and {\it NuSTAR}.
   We modeled the ultraviolet spectra to measure the strength and variability
   of the absorption, and used photoionization models to obtain its
   physical characteristics.
   }
   {We find new components of broad, blue-shifted absorption associated with
   Ly$\alpha$, \ion{N}{v}, \ion{Si}{iv}, and \ion{C}{iv} in our COS spectra.
   The absorption extends
   from velocities near zero in the rest-frame of the host galaxy to $-6200$
   $\rm km~s^{-1}$.
   These features appear for the first time in NGC 3783
   at the same time as heavy soft X-ray absorption seen in the {\it XMM-Newton}
   X-ray spectra. The X-ray absorption has a column density of
   $\sim 10^{23}~\rm cm^{-2}$, and it partially covers the X-ray continuum source.
   Combining the X-ray column densities with the UV spectral observations
   yields an ionization parameter for the obscuring gas of
   log $\xi =1.84^{+0.4}_{-0.2}$ $\rm erg~cm~s^{-1}$.
   Despite the high intensity of the UV continuum in NGC 3783,
   F(1470 \AA)=$8 \times 10^{-14}~\rm erg~cm^{-2}~s^{-1}~\AA^{-1}$,
   the well known narrow UV absorption lines are deeper than in earlier
   observations in unobscured states, and low ionization
   states such as \ion{C}{iii} appear, indicating that the
   narrow-line gas is more distant from the nucleus and is being shadowed by
   the gas producing the obscuration.
   Despite the high continuum flux levels in our observations of NGC 3783,
   moderate velocities in the UV broad line profiles have substantially diminished.
}
   {
   We suggest that a collapse of the broad line region has led to the outburst
   and triggered the obscuring event.
   }


\authorrunning{G. A. Kriss et al.}

\titlerunning{HST/COS observations of the obscuring outflow in NGC 3783}

   \maketitle
%

\section{Introduction}

Outflows from active galactic nuclei (AGN) may be the regulating mechanism that
links the growth of supermassive black holes at galaxy centers to the size
of the host galaxy. 
A possible outcome of such a linkage is the correlation between central
velocity dispersions in galaxies and the masses of their central black holes
\citep{Magorrian98, Ferrarese00, Gebhardt00, Kormendy13}.
Feedback from outflows may also regulate the overall mass and size of
the host galaxy
\citep{Silk98,King03,Ostriker10,Soker10,Faucher12,Zubovas14,Thompson15}.

AGN outflows manifest themselves in a variety of forms, from narrowly
collimated radio jets to broad, wide-spread winds. In the latter case, these
winds are often identified via their broad, blue-shifted absorption features
in X-ray and ultraviolet spectra \citep{Crenshaw03},
or extended, red and blue-shifted emission-line regions
\citep{Liu13a, Liu13b, Liu14}.
Again, the mechanisms for these various manifestations may vary, from
radiatively driven \citep{Murray95, Murray97, Proga00, Thompson15} or
magnetically accelerated \citep{Konigl94, Fukumura10} winds originating from the
accretion disk,
or thermal winds originating either from the accretion disk,
the broad-line region, or the obscuring torus \citep{KK95, KK01}.

Understanding the physical properties of outflows to ascertain how they work
is crucial for being able to model the interaction of central black holes with
their host galaxies. Observationally, in the X-ray and the UV, outflows
have appeared with a variety of characteristics, perhaps indicating several
mechanisms may be at work. Examples include the X-ray warm absorbers and
associated narrow UV absorption lines described by \citet{Crenshaw03};
ultra-fast outflows, typically only visible as broad, highly blue-shifted
Fe K features \citep{Pounds03,Reeves09,Tombesi10,Nardini15}; and the newly discovered
obscuring outflows showing strong soft X-ray absorption accompanied by
broad, fast, blue-shifted UV absorption lines: NGC~5548 \citep{Kaastra14},
Mrk~335 \citep{Longinotti13}, NGC~985 \citep{Ebrero16}, and most recently,
NGC~3783 \citep{Mehdipour17}.

In the case of obscuring outflows, the gas appears to be mildly ionized and
has high column density ($10^{22}-10^{23}~\rm cm^{-2}$). This produces strong
soft X-ray absorption, but no visible spectral features that allow diagnostics
of the kinematics or ionization state of the gas.
The crucial element in all the cases cited above is the availability of 
contemporaneous UV spectra. The UV absorption lines that appear in these
events provide the necessary diagnostics that show gas outflowing (blue-shifted)
with velocities and ionization states consistent with an origin in, or interior
to the broad-line region (BLR). 
With no UV spectra, such obscuration events would be indistinguishable from
other X-ray eclipsing events as studied by \citet{Markowitz14}, which could be
caused just as easily by clouds in transverse motion as 
opposed to having a significant outflow component.

To understand the nature of obscuring outflows better and study their potential
relationship to X-ray eclipsing events, we undertook a monitoring program
with {\it Swift} \citep{Gehrels04} to find potential obscuring events that we
could then study in detail with multiwavelength observations using
{\it XMM-Newton} \citep{Jansen01}, {\it NuSTAR} \citep{Harrison13},
and the Cosmic Origins Spectrograph (COS) on the
{\it Hubble Space Telescope (HST)}.
\cite{Mehdipour17} presented preliminary results from this campaign.
They found that the obscuring gas has kinematics and physical characteristics
comparable to gas normally associated with the BLR.
Thus, further study of obscuring outflows may offer some insights into the
physical structure of the BLR.
Reverberation mapping \citep{Blandford82, Peterson93} of the BLR has
indicated that motions are consistent with Keplerian motion in a
gravitational field dominated by the central black hole
\citep{Krolik91, Peterson99}.
Recent advances allowing two-dimensional reverberation mapping
in both spatial and velocity dimensions
\citep{Horne04, Bentz10, Grier13, Pancoast14a, Pancoast14b}
confirm the dominance of Keplerian motions, but also show evidence for
inflows and outflows associated with the BLR.
Models of accretion-disk winds have often suggested that the BLR may be an
observational manifestation of such winds, either
radiatively driven by line opacity
\citep{Murray95, Murray97, Proga00, Thompson15} or
by radiation pressure on dust \citep{Czerny11, Czerny17, Baskin18}, or
magnetohydrodynamic \citep{Konigl94, Fukumura10}.
If obscuring outflows are related to such winds that produce the BLR, then
their transient nature and its possible relationship to changes in the BLR
may provide additional insights into how the BLR forms and evolves.

NGC 3783 has been studied extensively in past UV and X-ray observational
campaigns.
The reverberation mapping campaign using the
{\it International Ultraviolet Explorer} (IUE) and ground-based observatories
in 1991--1992 \citep{Reichert94, Stirpe94} established the size of the
broad-line region (BLR) at 4--10 lt-days based on the lags of prominent
emission lines (Ly$\alpha$, \ion{C}{iv}, \ion{Mg}{ii}, H$\beta$) relative to
variations in the continuum emission.
In 2000--2001 an intensive X-ray and UV monitoring campaign obtained many
observations of NGC~3783 using {\it Chandra} \citep{Kaspi02},
{\it HST} \citep{Gabel03}, and the
{\it Far Ultraviolet Spectroscopic Explorer (FUSE)} \citep{Gabel03}.
The high-resolution X-ray spectra revealed details of the X-ray warm
absorber \citep{Kaspi02, Netzer03} and its relationship to the narrow
intrinsic absorption lines \citep{Gabel03, Gabel03b, Gabel05b}.
The intrinsic UV absorption lines comprised four discrete components
at outflow velocities ranging from $-$1352 to $-539~\rm km~s^{-1}$,
and the ensemble closely matched the kinematic appearance of the X-ray
absorption lines in the {\it Chandra} spectra \citep{Gabel03}.
Their absorption depth displayed variations consistent with a
photoionization response to changes in the UV continuum, and the
density-sensitive \ion{C}{iii}* $\lambda1176$ multiplet yielded a
density of log $\rm n_e = 4.5~cm^{-3}$ for Component \#1
($\rm v = -1311~km~s^{-1}$),
implying a distance of 25 pc for the gas producing the narrow UV
absorption lines \citep{Gabel05b}.
In contrast to the narrow UV absorption lines common in other Seyfert
galaxies \citep{Crenshaw03},
Component \#1 has appeared to ``decelerate", with its centroid evolving
from an outflow velocity of $-1352~\rm km~s^{-1}$ to $-1043~\rm km~s^{-1}$
\citep{Scott14} over 14 years.
This unusual kinematic behavior makes NGC 3783 an interesting object
for continuing studies.

In this paper we describe the HST/COS observations and their analysis that were
part of the detection of a new obscuring outflow in NGC~3783 by
\cite{Mehdipour17}.
In \S2 we describe the UV and optical observations and our data reduction methods.
In \S3 we model the UV and optical spectra and present an analysis of the
physical properties of the obscuring outflow and the evolution of the narrow
intrinsic UV absorption lines over the past 15 years.
\S4 discusses the implications of our observations for the structure of the BLR
and for the origin of the
obscuring outflow and its potential influence on the host galaxy.
\S5 presents our conclusions.


\section{Observations and Data Reduction}

During {\it Swift} Cycle 12 in 2016 November,
we detected spectral hardening in NGC~3783 indicative of an obscuring event.
We triggered coordinated {\it XMM-Newton, NuSTAR}, and HST/COS
observations on 2016-12-12 (Visit 3 in Program 14481) and 2016-12-21 (Visit 4).
In addition, on 2016-12-12 we obtained ground-based spectroscopy of the
H$\beta$ region using the Fiber-fed Extended Range Optical Spectrograph
(FEROS) on the Max Planck Gesellschaft/European Southern Observatory
(MPG/ESO) 2.2-m telescope.
For a baseline spectral comparison to NGC~3783 in the unobscured state,
we use archival {\it HST} UV and optical spectra obtained with the
Space Telescope Imaging Spectrograph (STIS) \citep{Woodgate98}.

\subsection{COS Observations}
Each COS observation consisted of a two-orbit visit using gratings G130M and
G160M to cover the 1130--1800 \AA\ wavelength range at a resolving
power of $\sim$15,000 \citep{Green12}.
We used multiple central wavelength settings and multiple FP-POS positions
to cover the gaps between detectors A and B for each grating, and to sample
the spectrum on different sections of each detector to allow for removal of
detector artifacts and other flat-field anomalies.
Table \ref{tab:cos_obs} provides a summary of the various COS exposures.

\begin{figure*}[!tbp]
  \centering
   \includegraphics[width=17cm, angle=-90,, scale=0.45, trim=0 0 255 0]{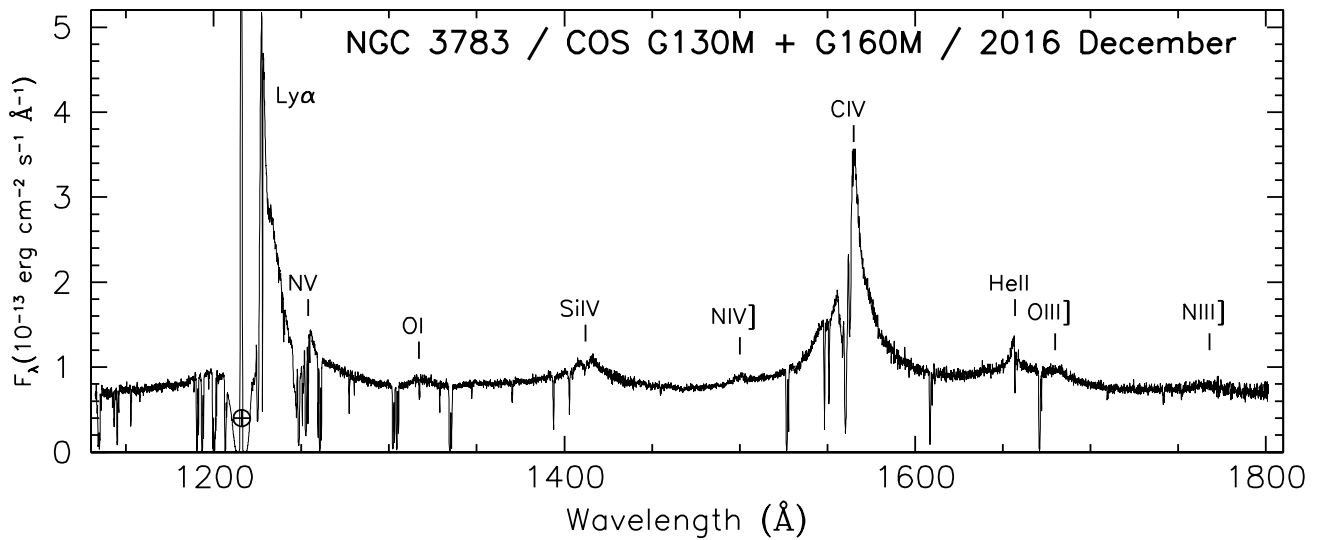}
  \vskip 15pt 
  \caption{Calibrated and merged COS spectrum of NGC~3783 from 2016 December.
Data are binned by 8 pixels, or approximately one resolution element.
We label the most prominent emission features. 
Geocoronal emission in the center of the Milky Way
Ly$\alpha$ absorption trough is indicated with an Earth symbol.}
  \label{fig_cosfull}
\end{figure*}

Using updated wavelength calibrations, flat fields, and flux calibrations,
we reprocessed our individual exposures
as described by \citet{Kriss11b} and \cite{DeRosa15}.
We then cross-correlated each exposure with prior HST/STIS observations of
NGC~3783 \citep{Gabel03} to adjust the zero-points of their wavelength scales
before combining them into merged spectra representing each visit, and the
two 2016 visits combined. Figure \ref{fig_cosfull} shows the full merged
spectrum from both 2016 visits.
Since NGC~3783 was exceptionally bright during both of our visits
(mean flux F(1470 \AA) = $7.6 \times 10^{-14}~\rm erg~cm^{-2}~s^{-1}~\AA^{-1}$
compared to the historical average of
$5.0 \times 10^{-14}~\rm erg~cm^{-2}~s^{-1}~\AA^{-1}$ \citep{Dunn06}),
each merged spectrum achieves a signal-to-noise (S/N) ratio
exceeding 40 per resolution
element over the wavelength range from Ly$\alpha$ to the \ion{C}{iv} line.

\begin{table*}
\begin{center}
        \caption[]{UV and Optical Observations of NGC~3783}
        \label{tab:cos_obs}
\begin{tabular}{l c c c c c}
\hline\hline
Data Set Name & Date & Start Time & Start Time & Exposure Time & Grating/Tilt/FP-POS\\
              &      &   (GMT)    &    (MJD)   & (s)           & \\
\hline
obgu03010 & 2011-03-23 & 19:41:12 & 55643.820286 & $\phantom{0}$696 & STIS/CCD/G430M \\
obgu03020 & 2011-03-23 & 19:53:58 & 55643.829152 & $\phantom{0}$696 & STIS/CCD/G430M \\
obgu03030 & 2011-03-23 & 21:11:46 & 55643.883179 & $\phantom{0}$696 & STIS/CCD/G430M \\
lbgu19010 & 2011-05-26 & 10:40:51 & 55707.445041 & $\phantom{0}$481 & G130M/1291/3 \\
lbgu19020 & 2011-05-26 & 10:52:19 & 55707.452998 & $\phantom{0}$481 & G130M/1300/3 \\
lbgu19030 & 2011-05-26 & 11:03:47 & 55707.460961 & $\phantom{0}$481 & G130M/1309/3 \\
lbgu19040 & 2011-05-26 & 12:05:01 & 55707.503484 & $\phantom{0}$481 & G130M/1318/3 \\
lbgu19050 & 2011-05-26 & 12:17:11 & 55707.511933 & $\phantom{0}$2425 & G160M/1589/3 \\
lbgu19060 & 2011-05-26 & 12:30:34 & 55707.521227 & $\phantom{0}$2425 & G160M/1600/3 \\
lbgu19070 & 2011-05-26 & 13:46:04 & 55707.573657 & $\phantom{0}$2425 & G160M/1611/3 \\
lbgu19080 & 2011-05-26 & 13:59:27 & 55707.582951 & $\phantom{0}$2425 & G160M/1623/3 \\
lc3x01010 & 2013-03-30 & 14:48:30 & 56381.617014 & $\phantom{0}$460 & G130M/1291/3 \\
lc3x01020 & 2013-03-30 & 14:59:23 & 56381.624572 & $\phantom{0}$460 & G130M/1300/3 \\
lc3x01030 & 2013-03-30 & 15:52:17 & 56381.661308 & $\phantom{0}$460 & G130M/1309/3 \\
lc3x01040 & 2013-03-30 & 16:03:10 & 56381.668866 & $\phantom{0}$340 & G130M/1318/3 \\
lc3x01050 & 2013-03-30 & 16:13:24 & 56381.675983 & $\phantom{0}$580 & G160M/1589/3 \\
lc3x01060 & 2013-03-30 & 16:26:08 & 56381.684815 & $\phantom{0}$580 & G160M/1600/3 \\
lc3x01070 & 2013-03-30 & 17:28:05 & 56381.727836 & $\phantom{0}$580 & G160M/1611/3 \\
lc3x01080 & 2013-03-30 & 17:40:49 & 56381.736678 & $\phantom{0}$424 & G160M/1623/3 \\
FEROS     & 2016-12-12 & 07:36:00 & 57734.316667 & 1800 & FEROS \\
$\rm ld3e03kgq^{\mathrm{a}}$ & 2016-12-12 & 12:13:56 & 57734.509676 & $\phantom{0}$475 & G130M/1291/3 \\
ld3e03kkq & 2016-12-12 & 12:23:55 & 57734.516609 & $\phantom{0}$475 & G130M/1291/4 \\
ld3e03kmq & 2016-12-12 & 12:35:16 & 57734.524502 & $\phantom{0}$455 & G130M/1327/1 \\
ld3e03koq & 2016-12-12 & 12:44:56 & 57734.531204 & $\phantom{0}$455 & G130M/1327/2 \\
ld3e03kyq & 2016-12-12 & 13:38:36 & 57734.568472 & $\phantom{0}$564 & G160M/1600/3 \\
ld3e03l2q & 2016-12-12 & 13:50:04 & 57734.576436 & $\phantom{0}$564 & G160M/1600/4 \\
ld3e03l4q & 2016-12-12 & 14:02:48 & 57734.585278 & $\phantom{0}$564 & G160M/1623/1 \\
ld3e03l6q & 2016-12-12 & 14:14:16 & 57734.593241 & $\phantom{0}$564 & G160M/1623/2 \\
$\rm ld3e04pmq^{\mathrm{a}}$ & 2016-12-21 & 15:25:00 & 57743.642361 & $\phantom{0}$475 & G130M/1291/3 \\
ld3e04poq & 2016-12-21 & 15:34:59 & 57743.649294 & $\phantom{0}$475 & G130M/1291/4 \\
ld3e04pqq & 2016-12-21 & 15:46:20 & 57743.657187 & $\phantom{0}$455 & G130M/1327/1 \\
ld3e04psq & 2016-12-21 & 15:56:00 & 57743.663889 & $\phantom{0}$455 & G130M/1327/2 \\
ld3e04puq & 2016-12-21 & 16:07:43 & 57743.672026 & $\phantom{0}$564 & G160M/1600/3 \\
ld3e04pwq & 2016-12-21 & 17:00:22 & 57743.708588 & $\phantom{0}$564 & G160M/1600/4 \\
ld3e04pyq & 2016-12-21 & 17:13:06 & 57743.717431 & $\phantom{0}$564 & G160M/1623/1 \\
ld3e04q0q & 2016-12-21 & 17:24:34 & 57743.725394 & $\phantom{0}$564 & G160M/1623/2 \\
\hline
\end{tabular}
\end{center}
{\bf Notes.}\\
$^{\mathrm{a}}$ We collectively refer to the visits on 2016-12-12 as Visit 3, or ``v3".\\
$^{\mathrm{b}}$ We collectively refer to the visits on 2016-12-21 as Visit 4, or ``v4".\\

\end{table*}

\subsection{FEROS Observations}

We observed NGC 3783 with the Fibre-fed Extended Range Optical Spectrograph
\citep[FEROS]{Kaufer99} at the MPG/ESO 2.2-m telescope at the ESO La Silla
Observatory.
Our spectra covered the wavelength range 3500--9200 \AA\ at a resolving
power of 48,000.
Starting at 07:36 GMT on 2016-12-12, three 10 minute exposures
were performed successively. Wavelength calibration frames were obtained
with ThArNe lamps.

The FEROS spectra were reduced manually using the standard pipeline
(FEROS-DRS\footnote{
https://www.eso.org/sci/facilities/lasilla/instruments/feros/tools/drs.html})
based on ESO-MIDAS.
Before the actual data reduction, the three exposures were averaged
and a cosmic ray rejection was performed. The successive data reduction included
flat fielding, bias and background subtraction, and wavelength calibration.

\subsection{Archival HST Data}

In 2000--2001 an intensive X-ray and UV monitoring campaign obtained many
observations of NGC~3783 using {\it Chandra} \citep{Kaspi02},
{\it HST} \citep{Gabel03}, and the
{\it Far Ultraviolet Spectroscopic Explorer (FUSE)} \citep{Gabel03}.
We obtained the calibrated STIS spectra from the Mikulski Archive for Space
Telescopes (MAST) and combined them as an unweighted average to use as our
baseline for comparison of the new COS spectra.

In 2011, HST program 12212 (PI: M. Crenshaw) obtained an optical long-slit
spectrum of the nuclear region of NGC~3783 using the STIS CCD, grating G430M,
and the 52$\times$0.2$\arcsec$ slit. This spectrum covers the wavelength
region surrounding H$\beta$ and [\ion{O}{iii}], and we use it as our baseline
for comparing our ground-based spectrum in the obscured state to an
unobscured state.
Table \ref{tab:cos_obs} gives the observational details for these spectra.
The STIS spectra were taken at three dithered slit positions to
facilitate removal of cosmic rays and defective pixels.
After correcting the data for charge-transfer inefficiency using a pixel-based
algorithm based on \cite{Anderson10}, we shifted the
calibrated two-dimensional images to a common geometric position using
integer-pixel shifts and combined them using a median filter to reject
bad pixels. After this process, some cosmic ray residuals were still present.
We identified these interactively, and interpolated across them using the
fluxes in the adjacent pixels.

\section{Data Analysis}

The absorption in the blue wings of the Ly$\alpha$, \ion{N}{v}, \ion{Si}{iv},
and \ion{C}{iv} emission lines in NGC~3783 is difficult to discern on the
scale of Figure \ref{fig_cosfull}, but it is more readily apparent in
Figure 3 of \cite{Mehdipour17}, where one can compare it directly to the
high-quality STIS spectrum obtained in 2001 \citep{Gabel03}
during an unobscured epoch.
To quantitatively assess the properties of the obscuration that appears in the
COS spectrum, we start with a model of the emission lines and continuum based on
the 2001 STIS spectrum. Our model is similar to the comprehensive model
used for NGC 5548 \citep{Kaastra14}
and consists of a power law continuum,
$\rm F_\lambda = F_{1000\AA} (\lambda / 1000 \AA)^{-\alpha}$,
that is reddened by fixed extinction of E(B-V)=0.107
\citep{Schlafly11}, and Gaussian emission components for each significant
emission line.
As usual, the bright components require several Gaussians.
Ly$\alpha$  and \ion{C}{iv} are each comprised
of 4 Gaussians: a narrow component with full-width at half-maximum
$\rm (FWHM) \sim 900 \rm~km~s^{-1}$,
a medium-broad component $\rm (FWHM\sim 2500~km~s^{-1}$),
a broad component $\rm (FWHM\sim 4500~km~s^{-1}$), and a very broad component
$\rm (FWHM\sim 10,000~ km~s^{-1}$).
\ion{N}{v}, \ion{Si}{iv} and \ion{He}{ii} require only three components
(narrow, medium broad or broad, and very broad),
and the weaker lines usually require only a broad component, or
a narrow plus a medium-broad to broad.
For \ion{N}{v}, \ion{Si}{iv}, and \ion{C}{iv}, we allow for both lines in the
doublets for the narrow, medium-broad, and broad components,
but only assume one component for the very broad line.
We also assume the doublets are optically thick and have flux ratios of 1:1.
The whole emission model is then absorbed by foreground Galactic \ion{H}{i}
Ly$\alpha$ as a damped Lorentzian profile with column density
N(HI)=$9.59 \times 10^{20}~\rm cm^{-2}$ \citep{Murphy96}.

We note that this Gaussian decomposition is a semi-empirical model and that the
individual components do not necessarily represent physically distinct portions
of the line-emitting regions. They also are neither independent nor orthogonal.
However, in both NGC 5548 and NGC 3783, there are distinct narrow portions of
the line profile that are stable over time in both width, velocity, and flux
\citep{Crenshaw09}.
As we show later, the narrow-line components in the spectra
at all epochs vary by $\ltsim 10$\%.
The lack of variability implies that these narrow components are likely
representative of emission from
the narrow-line region at some distance ($\sim$1 pc) from the nucleus.
This is important since these narrow components are probably not absorbed by the
obscurer.

We optimize parameters for each component in the model using the spectral
fitting program {\tt specfit} \citep{Kriss94} in IRAF.
The wavelength regions for our fits exclude all absorption lines,
both intrinsic to NGC~3783, and the foreground lines from the interstellar
medium (ISM). Since the intrinsic absorbers affect significant portions of the
blue sides of all the lines, best-fit parameters are largely
(but not exclusively) determined by the red sides of the profiles and the line
wings ($>1500~\rm km~s^{-1}$ from line center for the STIS spectrum,
and $>6500~\rm km~s^{-1}$ for the 2016 COS spectra).
For each spectrum, we optimize our fit in stages.
The continuum normalization and power-law index are initially fixed using
continuum regions near the blue and red ends of the spectrum.
We initially fix the line centers at the host galaxy systemic velocity,
z=0.00973, a value determined via \ion{H}{i} 21-cm measurements
\citep{Theureau98}.
We interactively choose widths and fluxes for the lines that
fit well by eye.
Next, we let all the emission line fluxes vary. Once these have converged, we
let the continuum parameters also vary freely to optimize its shape
simultaneously with the broad wings of the emission lines.
At this point the fit bears a good resemblance to the actual spectrum, but
is not a good match in all details. To approach full convergence, we then
successively address the major emission-line regions independently, as
described below.
For these individual regions, we freeze the continuum parameters,
free the line widths and fluxes first, and finally the line centers.
Since the line centers, widths, and fluxes are free, we are not forcing any
assumptions of symmetry on the line profiles; slight asymmetries in the line
profiles can be accommodated by the differing centroids and widths of the
various components.
Once the major line-emitting regions have been fit
(Ly$\alpha +$\ion{N}{v}, \ion{Si}{iv}, \ion{C}{iv}, and \ion{He}{ii}),
we let all parameters vary freely to converge to a globally
optimized best fit.

In the sections below we separately describe in further detail our
fitting process for the STIS 2001 spectrum first, and then the COS
2016 spectra.

\subsection{Modeling the Unobscured STIS Spectrum of NGC 3783}

To model the COS spectra of NGC 3783, we establish a baseline model by fitting
the average STIS spectrum accumulated in the monitoring campaign from
2000--2001 that included STIS, {\it FUSE}, and
{\it Chandra} observations \citep{Gabel03, Kaspi02}.
This spectrum has high S/N, and modeling it is straightforward, especially for
\ion{C}{iv}, since much of the line profile is uncontaminated by absorption.
In this spectrum, after establishing preliminary continuum and line fluxes,
we then fit the \ion{C}{iv} profile in detail.
This is the least-blended, highest S/N ratio emission feature.
The complement of Gaussian widths and relative fluxes here serve as a starting
point for the other major features.
We then fit the Ly$\alpha +$\ion{N}{v} region,
followed by \ion{Si}{iv} and \ion{He}{ii}.
After all regions have been individually optimized, we iterate on
the final model of the full spectrum to obtain the best fit.
The best-fit emission-line properties are given in Table \ref{tab:stiscos_em}.

\begin{sidewaystable*}
\begin{center}
        \caption[]{Emission-Line Parameters for the STIS 2000--2001 and COS 2016 Spectra of NGC~3783}
        \label{tab:stiscos_em}
\begin{tabular}{l c c c c c c c c}
\hline\hline
        &      & \multicolumn{3}{c}{STIS 2000--2001} &   &  \multicolumn{3}{c}{COS 2016}   \\
Feature & $\rm \lambda_0^{\mathrm{a}}$ &  $\rm Flux^{\mathrm{b}}$ & $\rm v_{sys}^{\mathrm{c}}$ & $\rm FWHM^{\mathrm{d}}$ & \phantom{000} & $\rm Flux^{\mathrm{b}}$ & $\rm v_{sys}^{\mathrm{c}}$ & $\rm FWHM^{\mathrm{d}}$ \\
\ion{C}{iii}\* &   1176.01 & $   1.1\pm  0.4$ & $  -100 \pm  70$ & $ 1330 \pm  170$ & \phantom{000} & $  40.0\pm  4.4$ & $  150 \pm  20$ & $ 1510 \pm   50$\\
Ly$\alpha$ &   1215.67 & $  84.0\pm  2.7$ & $   -40 \pm  20$ & $  800 \pm   30$ & \phantom{000} & $ 100.0\pm  3.8$ & $   -40 \pm  20$ & $  800 \pm   20$\\
Ly$\alpha$ &   1215.67 & $ 240.0\pm  9.6$ & $    40 \pm  30$ & $ 2500 \pm   20$ & \phantom{000} & $ 270.0\pm  8.1$ & $    40 \pm  30$ & $ 2500 \pm   30$\\
Ly$\alpha$ &   1215.67 & $ 120.0\pm  4.6$ & $   670 \pm  50$ & $ 5550 \pm   40$ & \phantom{000} & $ 290.0\pm  9.7$ & $   850 \pm  30$ & $ 6330 \pm   30$\\
Ly$\alpha$ &   1215.67 & $ 300.0\pm  9.8$ & $  -270 \pm  30$ & $14940 \pm  100$ & \phantom{000} & $ 400.0\pm 13.0$ & $   150 \pm  40$ & $15960 \pm   60$\\
\ion{N}{v} blue &   1238.82 & $   6.4\pm  0.2$ & $   -40 \pm  20$ & $  980 \pm   20$ & \phantom{000} & $   7.6\pm  0.3$ & $   -40 \pm  30$ & $  980 \pm   60$\\
\ion{N}{v} red &   1242.80 & $   6.4\pm  0.2$ & $   -40 \pm  20$ & $  980 \pm   20$ & \phantom{000} & $   7.6\pm  0.3$ & $   -40 \pm  30$ & $  980 \pm   20$\\
\ion{N}{v} blue &   1238.82 & $  13.0\pm  0.7$ & $   350 \pm  40$ & $ 2850 \pm   40$ & \phantom{000} & $   1.6\pm  0.3$ & $   350 \pm  70$ & $ 2850 \pm  120$\\
\ion{N}{v} red &   1242.80 & $  13.0\pm  0.7$ & $   350 \pm  40$ & $ 2850 \pm   40$ & \phantom{000} & $   1.6\pm  0.3$ & $   350 \pm  70$ & $ 2850 \pm   120$\\
\ion{N}{v} &   1240.89 & $  40.0\pm  1.6$ & $   660 \pm  60$ & $10380 \pm  180$ & \phantom{000} & $  28.0\pm  2.9$ & $  1410 \pm 140$ & $12140 \pm  140$\\
\ion{Si}{ii} &   1260.42 & $   1.7\pm  0.1$ & $  -150 \pm  20$ & $ 1600 \pm   20$ & \phantom{000} & $   1.5\pm  0.2$ & $  -150 \pm  50$ & $ 1600 \pm   50$\\
\ion{O}{i}+\ion{Si}{ii} &   1304.46 & $  20.0\pm  0.8$ & $     0 \pm  30$ & $ 3460 \pm   40$ & \phantom{000} & $  11.0\pm  1.1$ & $   220 \pm  50$ & $ 2500 \pm   40$\\
\ion{C}{ii} &   1334.34 & $   9.6\pm  0.3$ & $     0 \pm  50$ & $ 3460 \pm   40$ & \phantom{000} & $   3.1\pm  0.1$ & $   110 \pm  40$ & $ 2500 \pm   20$\\
\ion{Si}{iv} blue &   1393.76 & $   8.3\pm  0.3$ & $   -90 \pm  20$ & $ 1640 \pm   30$ & \phantom{000} & $   4.3\pm  0.3$ & $  -280 \pm  50$ & $ 1170 \pm   70$\\
\ion{Si}{iv} red &   1402.77 & $   8.3\pm  0.3$ & $   -90 \pm  20$ & $ 1640 \pm   30$ & \phantom{000} & $   4.3\pm  0.3$ & $  -280 \pm  50$ & $ 1170 \pm   20$\\
\ion{Si}{iv} blue &   1393.76 & $  23.0\pm  0.7$ & $   120 \pm  20$ & $ 4460 \pm   30$ & \phantom{000} & $  23.0\pm  0.9$ & $  -270 \pm  30$ & $ 5160 \pm   60$\\
\ion{Si}{iv} red &   1402.77 & $  23.0\pm  0.7$ & $   120 \pm  20$ & $ 4460 \pm   30$ & \phantom{000} & $  23.0\pm  0.9$ & $  -270 \pm  30$ & $ 5160 \pm   20$\\
\ion{Si}{iv} &   1398.19 & $  63.0\pm  1.9$ & $ -1170 \pm  30$ & $12410 \pm   80$ & \phantom{000} & $  55.0\pm  2.3$ & $ -2730 \pm 40$ & $13480 \pm  290$\\
\ion{O}{iv}] &   1400.37 & $   5.2\pm  0.4$ & $     0 \pm  20$ & $ 1640 \pm   80$ & \phantom{000} & $   0.8\pm  0.1$ & $   900 \pm  60$ & $ 1170 \pm   20$\\
\ion{O}{iv}] &   1400.37 & ... &    ... &  ...  & \phantom{000} & $   9.1\pm  0.5$ & $   900 \pm  60$ & $ 5160 \pm   20$\\
\ion{N}{iv}] &   1485.80 & $  14.0\pm  0.5$ & $   -10 \pm  40$ & $ 2600 \pm   40$ & \phantom{000} & $   2.9\pm  0.4$ & $  -130 \pm  80$ & $ 1170 \pm   20$\\
\ion{N}{iv}] &   14z5.80 & ... &    ... &  ...  & \phantom{000} & $   8.3\pm  0.4$ & $  -130 \pm  80$ & $ 3250 \pm   20$\\
\ion{C}{iv} blue &   1548.19 & $  40.0\pm  1.7$ & $   -60 \pm  20$ & $  940 \pm   40$ & \phantom{000} & $  36.0\pm  1.7$ & $   -60 \pm  30$ & $  940 \pm   30$\\
\ion{C}{iv} red &   1550.77 & $  40.0\pm  1.7$ & $   -60 \pm  20$ & $  940 \pm   40$ & \phantom{000} & $  36.0\pm  1.7$ & $   -60 \pm  30$ & $  940 \pm   20$\\
\ion{C}{iv} blue &   1548.19 & $  80.0\pm  2.6$ & $   140 \pm  20$ & $ 2840 \pm   20$ & \phantom{000} & $   2.3\pm  0.6$ & $   140 \pm 20$ & $ 2840 \pm  20$\\
\ion{C}{iv} red &   1550.77 & $  80.0\pm  2.6$ & $   140 \pm  20$ & $ 2840 \pm   20$ & \phantom{000} & $   2.3\pm  0.6$ & $   140 \pm  20$ & $ 2840 \pm   20$\\
\ion{C}{iv} blue &   1548.19 & $  84.0\pm  2.7$ & $  -950 \pm  20$ & $ 4580 \pm   30$ & \phantom{000} & $ 180.0\pm  5.5$ & $  -710 \pm  30$ & $ 4640 \pm   30$\\
\ion{C}{iv} red &   1550.77 & $  84.0\pm  2.7$ & $  -950 \pm  20$ & $ 4580 \pm   30$ & \phantom{000} & $ 180.0\pm  5.5$ & $  -710 \pm  30$ & $ 4640 \pm   30$\\
\ion{C}{iv} &   1549.48 & $ 310.0\pm  9.3$ & $   -60 \pm  20$ & $10030 \pm   30$ & \phantom{000} & $ 340.0\pm 11.0$ & $   -70 \pm  30$ & $12710 \pm   40$\\
\ion{He}{ii} &   1640.45 & $  13.0\pm  0.5$ & $  -140 \pm  20$ & $  990 \pm   20$ & \phantom{000} & $  14.0\pm  0.6$ & $    10 \pm  50$ & $  820 \pm   30$\\
\ion{He}{ii} &   1640.45 & $   9.2\pm  0.3$ & $  -510 \pm  40$ & $ 3000 \pm  100$ & \phantom{000} & $   7.6\pm  1.2$ & $  1250 \pm  40$ & $ 3990 \pm  900$\\
\ion{He}{ii} &   1640.45 & $ 140.0\pm  4.4$ & $  -470 \pm  30$ & $11280 \pm   30$ & \phantom{000} & $ 160.0\pm  6.3$ & $  -120 \pm  30$ & $12460 \pm   50$\\
\ion{O}{iii}] &   1659.85 & $   3.6\pm  0.2$ & $     0 \pm  50$ & $ 1200 \pm  100$ & \phantom{000} & $   3.1\pm  0.4$ & $   150 \pm  40$ & $ 1210 \pm   30$\\
\ion{O}{iii}] &   1665.19 & $   7.6\pm  0.6$ & $     0 \pm  10$ & $ 1200 \pm  100$ & \phantom{000} & $   5.7\pm  0.4$ & $   150 \pm  20$ & $ 1210 \pm   20$\\
\ion{N}{iii}] &   1750.00 & ...  &    ... &  ...  & \phantom{000} & $  21.0\pm  0.7$ & $     0 \pm  30$ & $ 3270 \pm   40$\\
\hline
\end{tabular}
\end{center}
{\bf Notes.}\\
$^{\mathrm{a}}$ Vacuum rest wavelength of the spectral feature (\AA).\\
$^{\mathrm{b}}$ Integrated flux in units of $\rm 10^{-14}~erg~cm^{-2}~s^{-1}$.\\
$^{\mathrm{c}}$ Velocity (in $\rm km~s^{-1}$) relative to a systemic redshift of z = 0.00973 \citep{Theureau98}.\\
$^{\mathrm{d}}$ Full-width at half-maximum ($\rm km~s^{-1}$).\\
\end{sidewaystable*}

\subsubsection{Modeling the Unobscured C IV Profile}
\begin{figure}[!tbp]
\centering
\resizebox{1.0\hsize}{!}{\includegraphics[angle=270]{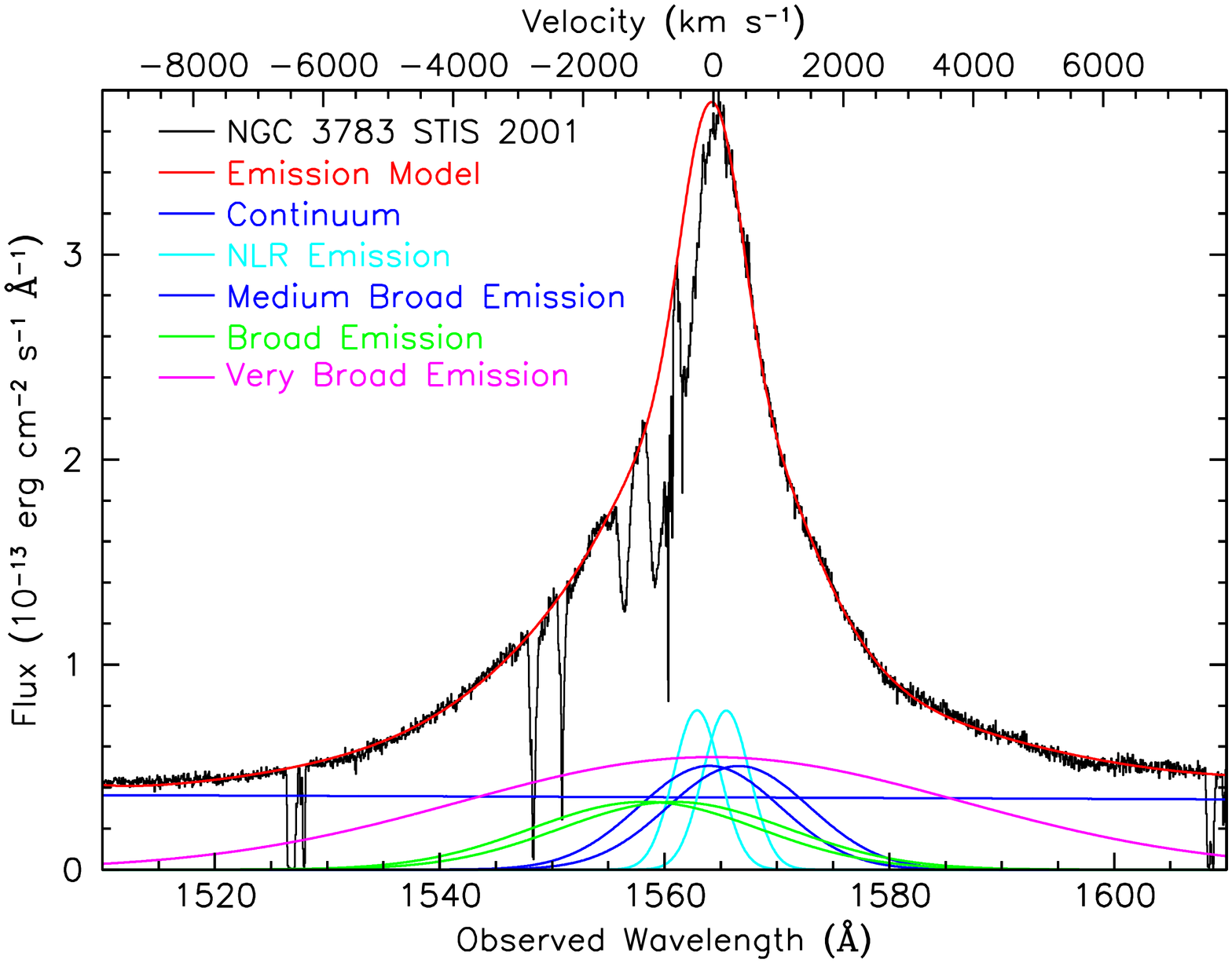}}
\caption{
Average STIS spectrum of the \ion{C}{iv} region from all observations in the
2000--2001 campaign re-binned into 0.05 \AA\ pixels (black histogram).
The solid red line tracing the data is the total emission model.
The key in the figure identifies the emission components in our model.
The velocity scale along the top axis is for the blue component of the
\ion{C}{iv} doublet, $\lambda1548.195$, relative to the host galaxy systemic
redshift, z=0.00973 \citep{Theureau98}.
}
\label{fig_stis_c4}
\end{figure}

Figure \ref{fig_stis_c4} shows the best fit, with the profiles of all the
\ion{C}{iv} emission components shown as well as the continuum level.
(Note that the best fit shown here is the final result following the global
optimization to the full spectrum after all regions have been separately fit.)
The \ion{C}{iv} narrow-line emission (cyan) is slightly blueshifted by
about $-265~\rm km~s^{-1}$.
The medium-broad emission (blue) and the very broad emission (magenta) lie near
the systemic velocity.
The broad components of the line profile (green) have a significant blue shift
of $-900~\rm km~s^{-1}$.

\subsubsection{Modeling the Unobscured Ly$\alpha$ Profile}

\begin{figure}[!tbp]
\centering
\resizebox{1.0\hsize}{!}{\includegraphics[angle=0]{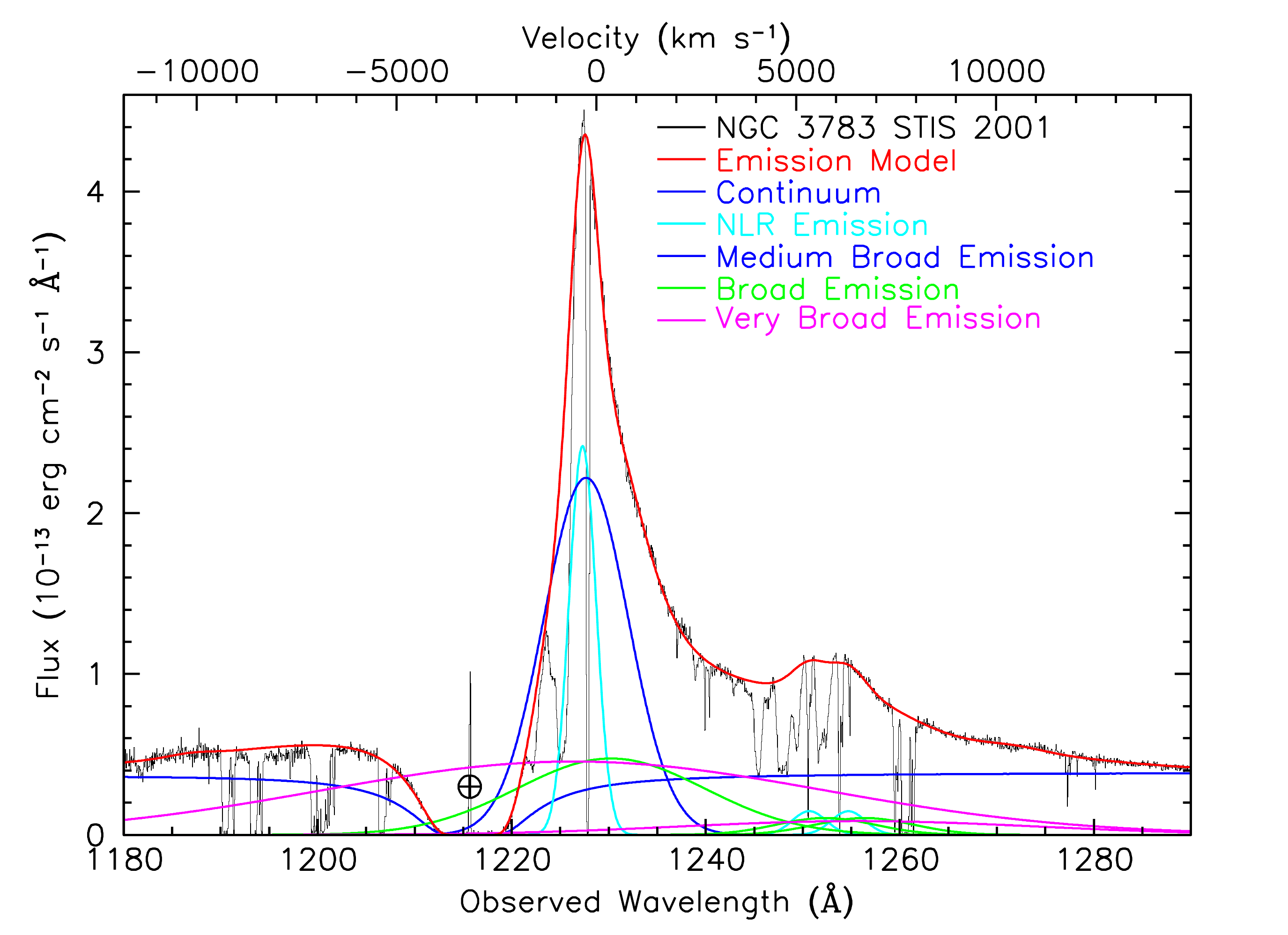}}
\caption{
Average STIS spectrum of the Ly$\alpha +$\ion{N}{v} region from all observations
in the 2000--2001 campaign re-binned into 0.05 \AA\ pixels (black histogram).
The solid red line tracing the data is the total emission model.
The key in the figure identifies the emission components in our model.
The velocity scale along the top axis is for 
Ly$\alpha$ $\lambda1215.67$, relative to the host galaxy systemic
redshift, z=0.00973 \citep{Theureau98}.
Geocoronal emission in the center of the Milky Way
Ly$\alpha$ absorption trough is indicated with an Earth symbol.
The solid blue line showing the continuum also illustrates the impact of the
damped Milky Way Ly$\alpha$ absorption. Note that the influence of this damping
profile extends all the way from 1180 \AA\ on through the peak of Ly$\alpha$
in NGC~3783 and into the \ion{N}{v} profile at 1250 \AA.
}
\label{fig_stis_lyan5}
\end{figure}

The Ly$\alpha +$\ion{N}{v} region is more complicated due to the blending of
the Ly$\alpha$ and \ion{N}{v} profiles and the
strong impact of damped Ly$\alpha$ absorption from the Milky Way.
Although we use the \ion{C}{iv} profile as a guide and choose initial values
for the Ly$\alpha$ components based on the \ion{C}{iv} best fit, we do not
constrain or tie these components to those determined for \ion{C}{iv}.
The final results do not wander far from these initial guesses, but we note that
trial attempts with different plausible initial conditions also result in good
fits that are slightly different; the blending and the foreground damped
absorption lead to fits that are highly degenerate.
Figure \ref{fig_stis_lyan5} illustrates the final fit to this region.
As for \ion{C}{iv}, the narrow-line emission is significantly blue shifted,
with a peak at $-340~\rm km~s^{-1}$.
The medium-broad emission is also blueshifted at $-250~\rm km~s^{-1}$.
In contrast to \ion{C}{iv}, the broad emission here is redshifted to
$+350~\rm km~s^{-1}$ (for both Ly$\alpha$ and \ion{N}{v}), but this may well be
biased because of the lack of a good view of the blue wing of Ly$\alpha$ and
the blending of \ion{N}{v} with Ly$\alpha$.

\subsubsection{Modeling the Unobscured Si IV Profile}

\begin{figure}[!tbp]
\centering
\resizebox{1.0\hsize}{!}{\includegraphics[angle=270]{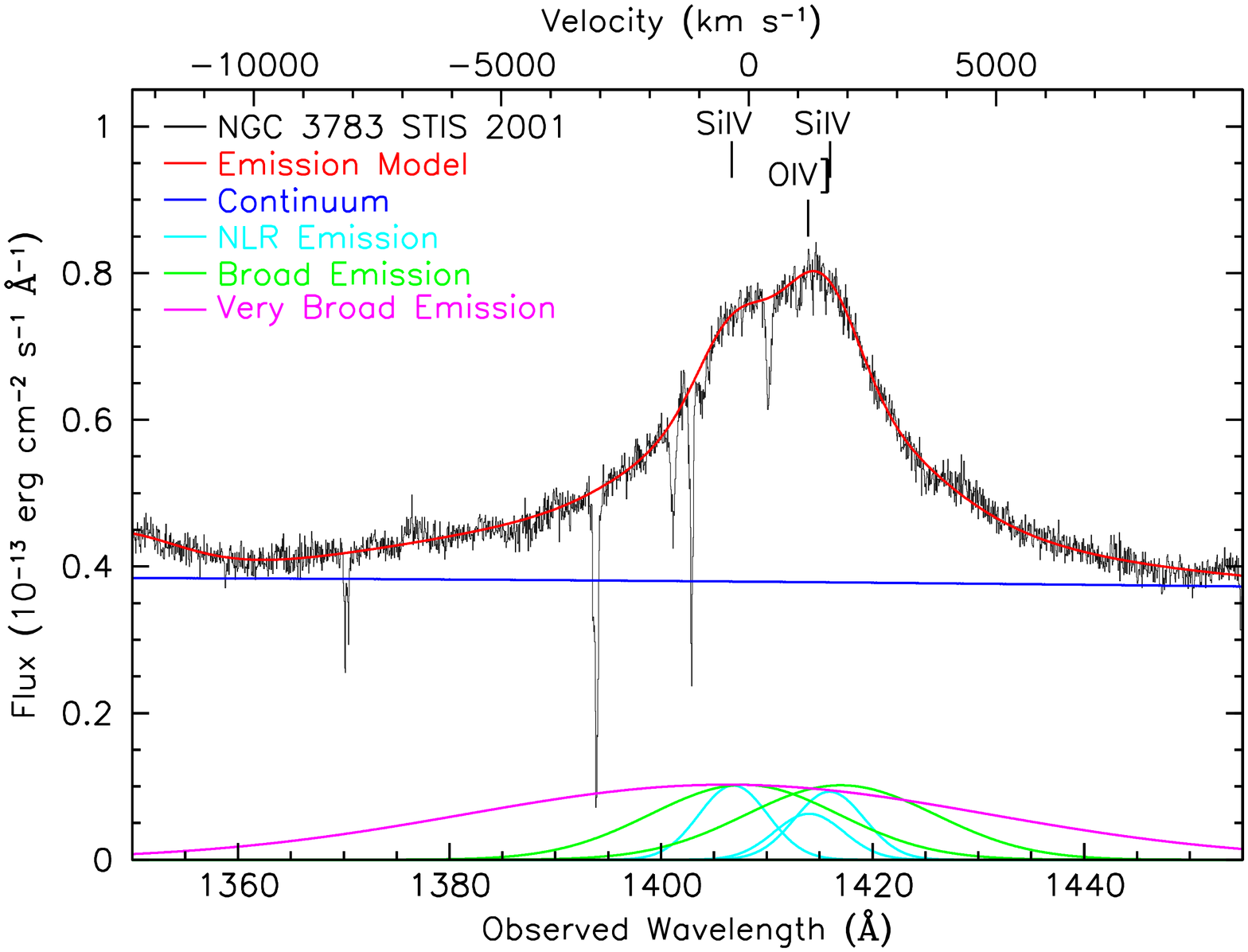}}
\caption{
Average STIS spectrum of the \ion{Si}{iv} region from all observations
in the 2000--2001 campaign re-binned into 0.05 \AA\ pixels (black histogram).
The solid red line tracing the data is the total emission model.
The key in the figure identifies the emission components in our model.
The velocity scale along the top axis is for 
\ion{Si}{iv} $\lambda1393.755$, relative to the host galaxy systemic
redshift, z=0.00973 \citep{Theureau98}.
}
\label{fig_stis_si4}
\end{figure}

Fitting the \ion{Si}{iv} region is simpler since there is less contaminating
intrinsic and foreground absorption. The line is fainter, however, and the
S/N is not as good as in the \ion{C}{iv} and Ly$\alpha$ regions.
Therefore an adequate fit requires only two narrow and two broad components
for the \ion{Si}{iv} doublets, and a single very broad \ion{Si}{iv} base.
\ion{Si}{iv} is also blended with several \ion{O}{iv}] transitions; however,
these appear to be weak, and only the cluster of the strongest transitions
at $\lambda$1401 is included in our model as an additional narrow component.
The best fit is shown in Figure \ref{fig_stis_si4}.
As with \ion{C}{iv} and Ly$\alpha$, the narrow lines are slightly blue shifted
at $-100~\rm km~s^{-1}$, and, similar to Ly$\alpha$, the broad lines are
slightly redshifted at $+120~\rm km~s^{-1}$.

\subsubsection{Modeling the Unobscured He II Profile}

\begin{figure}[!tbp]
\centering
\resizebox{1.0\hsize}{!}{\includegraphics[angle=270]{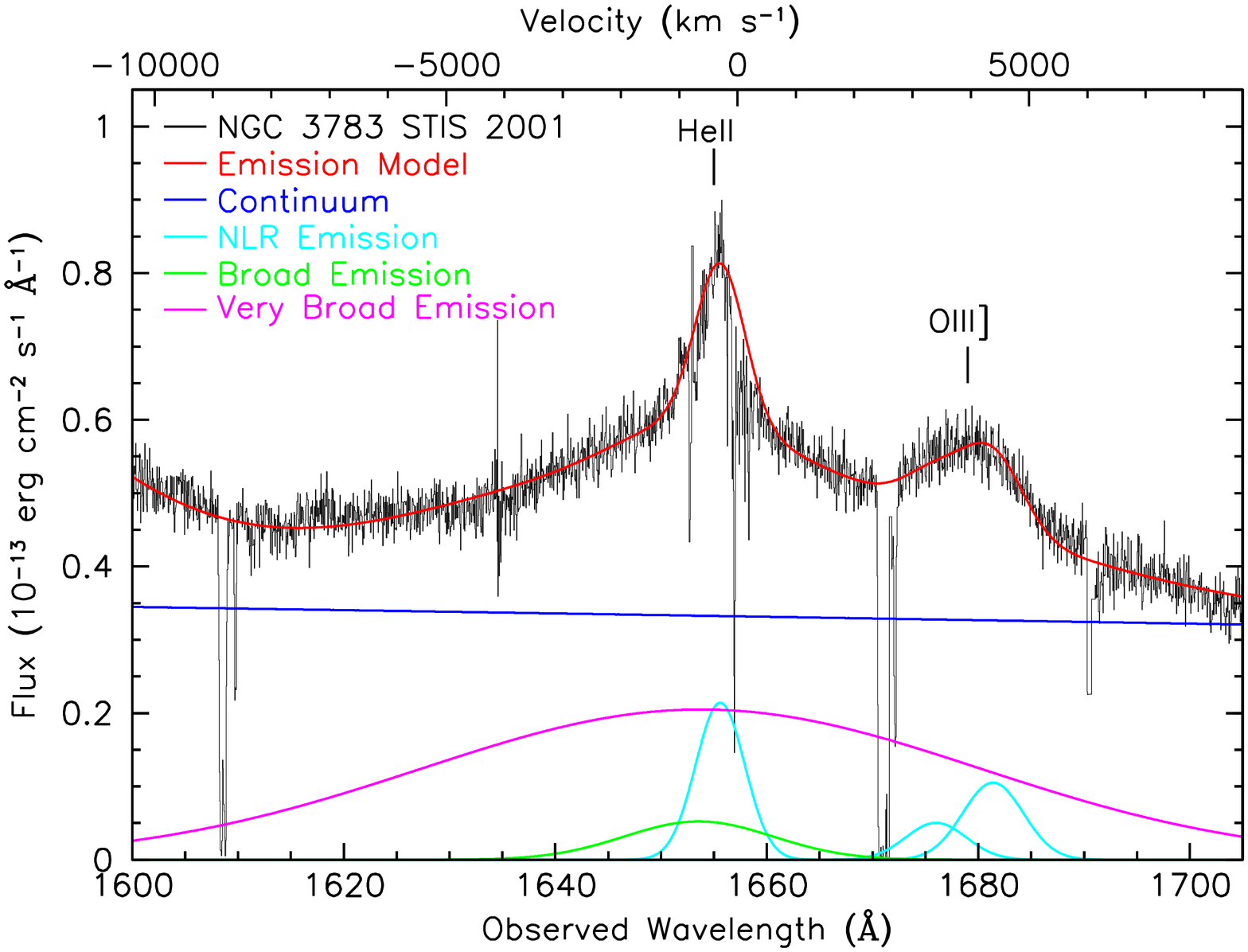}}
\caption{
Average STIS spectrum of the \ion{He}{ii} region from all observations
in the 2000--2001 campaign re-binned into 0.05 \AA\ pixels (black histogram).
The solid red line tracing the data is the total emission model.
The key in the figure identifies the emission components in our model.
The velocity scale along the top axis is for 
\ion{He}{ii} $\lambda1640.45$, relative to the host galaxy systemic
redshift, z=0.00973 \citep{Theureau98}.
All narrow absorption features are foreground ISM lines.
}
\label{fig_stis_he2}
\end{figure}

Finally, we show the fit to the \ion{He}{ii} region.
This is important for establishing a good template for
the broad-line region since \ion{He}{ii} $\lambda 1640$ is a recombination
line to the excited n=2 state, and therefore unlikely to be
affected by absorption.\footnote{
\ion{He}{ii} n=2 is 48 eV above the ground state, and maintaining a significant
population of ions in this state requires high temperatures (T $\gtsim 6 \times 10^5$ K), or high densities (n $\gtsim 10^{15}~\rm cm^{-3}$) neither of which
are typical of BLR conditions of T $\sim 10^4$ K and
n $\sim 10^{10}~\rm cm^{-3}$ \citep{Osterbrock}.
}
It is also similar in ionization to
\ion{C}{iv}, and should form in the same physical regions
and have similar overall kinematics.
Since \ion{He}{ii} emission is faint, like  \ion{Si}{iv},
only three components are required: a narrow, a broad, and a
very broad component. There is also slight contamination on
the red wing by two \ion{O}{iii}] transitions, which we
model with individual narrow components.
Also note that the red half of the line center is strongly absorbed by
foreground interstellar \ion{Al}{ii} $\lambda$1670.
Finally, this whole complex sits atop the far red wing of the
much brighter \ion{C}{iv} emission line.
Figure \ref{fig_stis_he2} shows the best fit to the
\ion{He}{ii} region. As for all previous emission lines,
the narrow-line component is slightly blue-shifted at
$-150~\rm km~s^{-1}$; the broad component is significantly
blue-shifted at $-520~\rm km~s^{-1}$, but not as strongly
as \ion{C}{iv}.

\subsection{Modeling the Average Obscured COS Spectrum of NGC 3783}

Now that we have a good baseline model for the emission spectrum of NGC 3783,
we can look at how the spectrum has changed from the unobscured state in
2001 to the obscured state that appears in our 2016 observation.
We first fit the unweighted average sum of the two spectra from 2016-12-12 and
2016-21-21.
As for the STIS spectrum, we fit the obscured COS selected regions at a time,
beginning with \ion{He}{ii}, which we do not expect to be absorbed, followed by
\ion{C}{iv}, Ly$\alpha+$\ion{N}{v}, and then \ion{Si}{iv}.
After all regions have been individually optimized, we iterate on
the final model of the full spectrum to obtain the best fit.
Table \ref{tab:stiscos_em} compares the best-fit emission-line properties
for the average obscured COS spectrum in 2016 to the STIS spectrum
from 2000--2001.
Once we have this fit to the average spectrum, we then scale the model to
adjust the continuum and line fluxes to the levels of the individual
observations on 2016-12-12 and 2016-12-21, and then optimize the fits to
each of these individual spectra.
Results for the two separate observations are given in
Table \ref{tab:cos3cos4_em}.
In the following sections, we describe the fits to each of the individual
regions in the obscured COS spectrum.

\subsubsection{Modeling the Obscured He II Profile}

\begin{figure}[!tbp]
\centering
\resizebox{1.0\hsize}{!}{\includegraphics[angle=270]{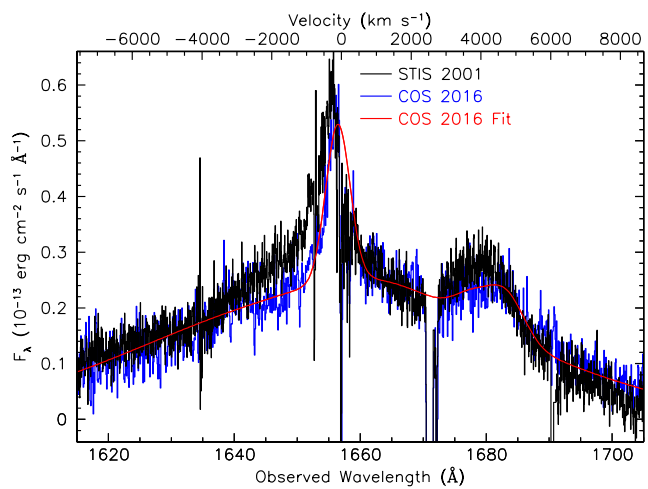}}
\caption{
Comparison of STIS and COS spectra of the region surrounding the \ion{He}{ii}
emission line.
The black histogram is the average STIS 2001 spectrum. The blue histogram is the
unweighted average COS spectrum from the 12 December 2016 and 21 December 2016
observations. Both spectra have the continuum subtracted as well as the
underlying emission
of the \ion{C}{iv} emission line. The STIS spectrum is scaled up
by $1.2\times$ to match the flux in the far wings of the COS spectrum.
The solid red line is the best-fit emission model for the COS spectrum.
The velocity scale along the top axis has zero velocity for \ion{He}{ii}
$\lambda1640.45$ at the host-galaxy redshift of 0.00973 \citep{Theureau98}.
All narrow absorption features are foreground ISM lines.
}
\label{fig_cos_he2}
\end{figure}

Since we do not expect the \ion{He}{ii} emission line to be
absorbed, we begin our fits with that line to see what intrinsic changes
in the profile of the broad emission lines may have taken place between
2001 and 2016. Figure \ref{fig_cos_he2} compares the STIS 2001 spectrum
to the COS 2016 spectrum. The continuum and the underlying emission of the
\ion{C}{iv} profile have been subtracted. The STIS spectrum is scaled up by a
factor of 1.2 to match the flux in the high-velocity wings.
\ion{O}{iii}] emission that sits on the red wing of \ion{He}{ii} has not
been removed.
One can see that overall, there is little change in the \ion{He}{ii}
profile. Interestingly, the narrow component has moved redward and now
lies at the systemic velocity.
To optimize the model to the COS spectrum, we adjust the line fluxes in
the model and the wavelength of the narrow component accordingly before
letting the fit iterate to a new minimum.
The resulting best fit is shown in Figure \ref{fig_cos_he2}.

\begin{figure}[!tbp]
\centering
\resizebox{1.0\hsize}{!}{\includegraphics[angle=270]{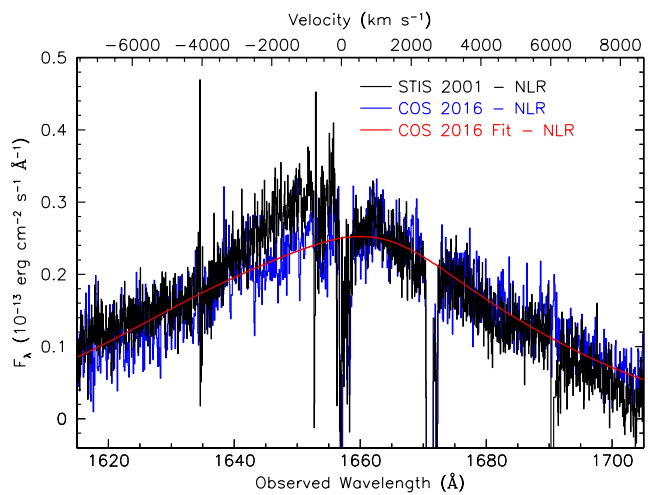}}
\caption{
Comparison of STIS and COS spectra of the region surrounding the \ion{He}{ii}
emission line omitting the narrow emission.
The black histogram is the average STIS 2001 spectrum with the narrow
components of \ion{He}{ii} and \ion{O}{iii}] subtracted.
The blue histogram is the
unweighted average COS spectrum from the 12 December 2016 and 21 December 2016
observations, also with the narrow components subtracted.
Both spectra have the continuum subtracted as well as the underlying emission
of the \ion{C}{iv} emission line. The STIS spectrum is scaled up
by $1.2\times$ to match the flux in the far wings of the COS spectrum.
The solid red line is the best-fit emission model for the COS spectrum.
The velocity scale along the top axis has zero velocity for \ion{He}{ii}
$\lambda1640.45$ at the host-galaxy redshift of 0.00973 \citep{Theureau98}.
}
\label{fig_cos_he2n}
\end{figure}

To compare the broad-line region profiles more closely, we subtract the
modeled narrow emission lines from each profile and show the result in
Figure \ref{fig_cos_he2n}.
Here you can see that the STIS and COS profiles are very similar, except
for a little added emission in the STIS spectrum on the blue side of line center.
Although the STIS spectrum exhibits this excess emission relative to COS,
there is no indication of broad absorption in the COS spectrum; the differences
between the two profiles are accommodated merely by adjusting the flux and wavelength
of the broad component (now shifted to the red as for the narrow component).
Table \ref{tab:stiscos_em} shows that the flux in the broad component of the
COS spectrum is 17\% lower compared to STIS, and shifted 1760 $\rm km~s^{-1}$
(10 \AA) to the red.
The fitted line profile passes smoothly through all the data in both line wings. 

\subsubsection{Modeling the Obscured C IV Profile}

\begin{figure}[!tbp]
\centering
\resizebox{1.0\hsize}{!}{\includegraphics[angle=270]{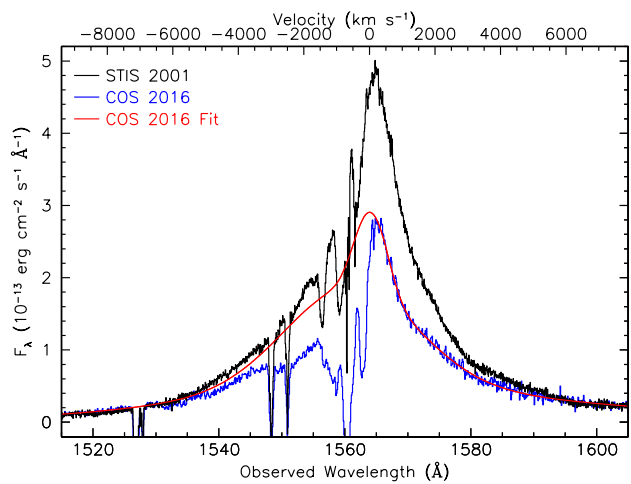}}
\caption{
Average STIS spectrum of the \ion{C}{iv} region from all observations
in the 2000--2001 campaign re-binned into 0.05 \AA\ pixels (black histogram).
The blue histogram is the average COS spectrum from 2016.
The STIS data have been scaled up by a factor of 1.45 so that the intensities
match in the broad wings.
The solid red line tracing the data is the total emission model for the
average 2016 COS spectrum.
The velocity scale along the top axis is for
\ion{C}{iv} $\lambda1548.195$, relative to the host galaxy systemic
redshift, z=0.00973 \citep{Theureau98}.
}
\label{fig_stiscosc4all}
\end{figure}

From our fit to the \ion{He}{ii} profile, we expect that our basic model for
 \ion{C}{iv} should also be very similar in the COS 2016 spectrum compared
to the STIS 2001 spectrum.
Figure \ref{fig_stiscosc4all} compares the \ion{C}{iv} profile from STIS 2001
to COS 2016. As for \ion{He}{ii}, both spectra have been continuum subtracted,
and the STIS spectrum has been scaled up by a factor of 1.45 so that the
intensities match in the broad wings beyond $\pm6000~\rm km~s^{-1}$.
For reference, for COS 2016 the continuum flux
F(1560\AA)=$7.5\times 10^{-14}~\rm erg~cm^{-2}~s^{-1}~\AA^{-1}$;
for STIS 2011, F(1560\AA)=$3.5\times 10^{-14}~\rm erg~cm^{-2}~s^{-1}~\AA^{-1}$.
Note that even though the continuum in 2016 is more than twice as bright as in
2001, the \ion{C}{iv} emission line is fainter, particularly in the mid section
of the profile. However, the narrow cores of each profile are nearly unchanged.
Indeed, in our best fit, the narrow-line component has the same velocity and width,
and is within 10\% of the flux observed in 2001.

\begin{figure}[!tbp]
\centering
\resizebox{1.0\hsize}{!}{\includegraphics[angle=270]{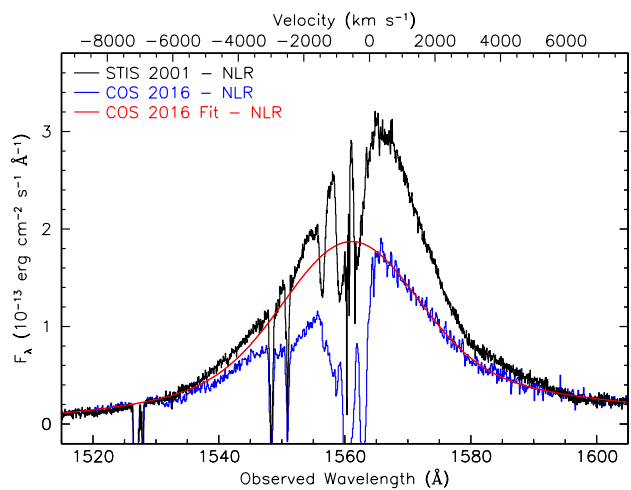}}
\caption{
Average STIS spectrum of the \ion{C}{iv} region from all observations
in the 2000--2001 campaign re-binned into 0.05 \AA\ pixels (black histogram)
with the narrow-line component of the fit subtracted from the data. 
The blue histogram is the average COS spectrum from 2016 with the
narrow-line component subtracted.
The STIS data have been scaled up by a factor of 1.45 so that the intensities
match in the broad wings.
The solid red line tracing the data is the total emission model for the
average 2016 COS spectrum with the narrow-line component removed.
The velocity scale along the top axis is for
\ion{C}{iv} $\lambda1548.195$, relative to the host galaxy systemic
redshift, z=0.00973 \citep{Theureau98}.
}
\label{fig_stiscosc4n}
\end{figure}

Given the apparently fixed intensity of the narrow-line emission,
Figure \ref{fig_stiscosc4n} compares the profiles if we subtract this fitted
emission component from the profiles of each spectrum, and from the best fit
curve. One can now see that most of the excess emission in the 2001 spectrum is
in what we identified as the medium-broad component in Figure \ref{fig_stis_c4},
which are the blue curves near a systemic velocity of zero in that figure.
Note that this component is not needed in our fit to either \ion{He}{ii}
emission profile.

\begin{figure}[!tbp]
\centering
\resizebox{1.0\hsize}{!}{\includegraphics[angle=270]{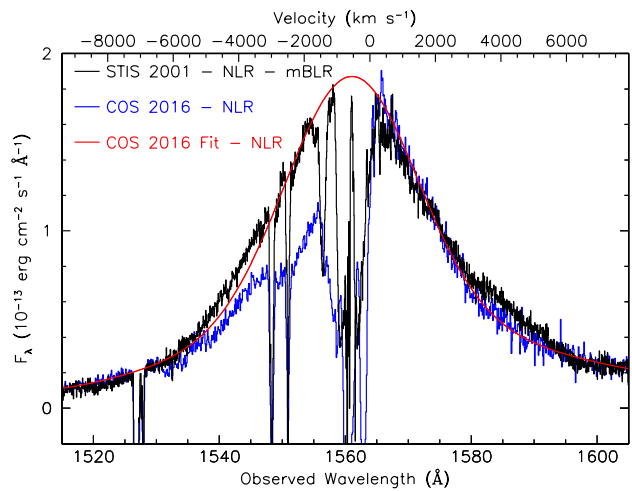}}
\caption{
Average STIS spectrum of the \ion{C}{iv} region from all observations
in the 2000--2001 campaign re-binned into 0.05 \AA\ pixels (black histogram)
with both the narrow-line and medium-broad components of the fit subtracted from
the data. 
The blue histogram is the average COS spectrum from 2016 with the
narrow-line and medium-broad components subtracted.
The STIS data have been scaled up by a factor of 1.45 so that the intensities
match in the broad wings.
The solid red line tracing the data is the total emission model for the
average 2016 COS spectrum with the narrow-line component removed.
The velocity scale along the top axis is for
\ion{C}{iv} $\lambda1548.195$, relative to the host galaxy systemic
redshift, z=0.00973 \citep{Theureau98}.
}
\label{fig_stiscosc4i}
\end{figure}

If we now remove this medium-broad component from the STIS spectrum and
compare the result to the observed profile in 2016, Figure \ref{fig_stiscosc4i}
shows that the shape of the two profiles match well in both the far wings of the
line (both red and blue), and on the red side of the line.
The minor differences in the profiles on the red side of line center are
accommodated by slight adjustments in the relative fluxes of the broad and
very broad components from 2001 to 2016.
However, there is a noticeable deficiency in flux on the blue side of the line
profile, which we interpret as absorption due to the appearance of the obscurer.

\begin{figure}[!tbp]
\centering
\resizebox{1.0\hsize}{!}{\includegraphics[angle=270]{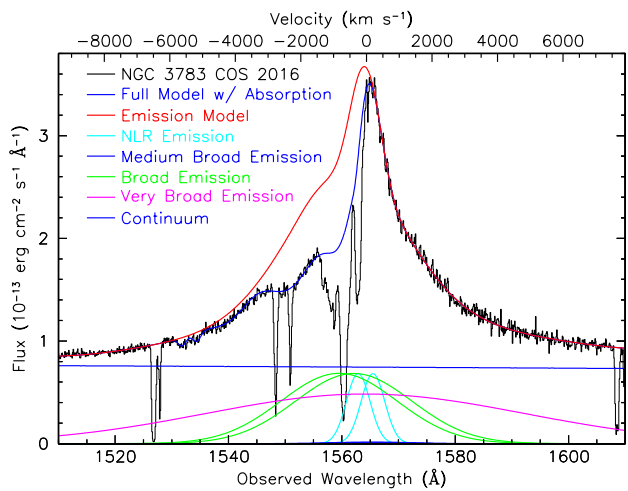}}
\caption{
Average COS spectrum of the \ion{C}{iv} region from observations in 2016
December (black histogram).
The solid red line tracing the data is the total emission model.
The solid blue line tracing the data is the emission model modified by
components that model the broad absorption.
The key in the figure identifies the emission components in our model.
The velocity scale along the top axis is for the blue component of the
\ion{C}{iv} doublet, $\lambda1548.195$, relative to the host galaxy systemic
redshift, z=0.00973 \citep{Theureau98}.
}
\label{fig_cos_c4}
\end{figure}

So, to adapt the \ion{C}{iv} profile from 2001 to what we observe in 2016,
we fixed the intensity of the narrow-line components at the 2001
value, scaled the intensity of the very broad emission up by the factor of
1.45 required to match the far wings, kept all line centers and widths fixed,
and allowed the medium-broad and broad fluxes to vary freely.
The result was that the medium-broad component's intensity dropped to nearly
zero, with most remaining flux in the broad component. We then freed all
parameters (all line centers, widths and intensities) and optimized the fit. 
Figure \ref{fig_cos_c4} shows the best-fit emission model and all its
components.

To provide a semi-empirical characterization of the blue-shifted absorption
in the \ion{C}{iv} profile, we added a series of \ion{C}{iv} absorption-line
doublets that coincide with the deepest inflection points in the blue wing
at outflow velocities of $-$2590, $-$4625, $-$5515, and $-$6025 $ \rm~km~s^{-1}$.
These doublets are modeled as Gaussians in optical depth with relative
velocities linked at the ratio of rest wavelengths, identical widths
of $\sim500 \rm~km~s^{-1}$ for each doublet component,
and blue to red optical depths assumed to have a 2:1 ratio.
This characterization enables us in our subsequent modeling to test whether
similar absorption is present in other lines such as Ly$\alpha$, \ion{N}{v},
and \ion{Si}{iv}.
The full final model including these additional absorption components is also
shown in Figure \ref{fig_cos_c4}.

\begin{figure}[!tbp]
\centering
\resizebox{1.0\hsize}{!}{\includegraphics[angle=270]{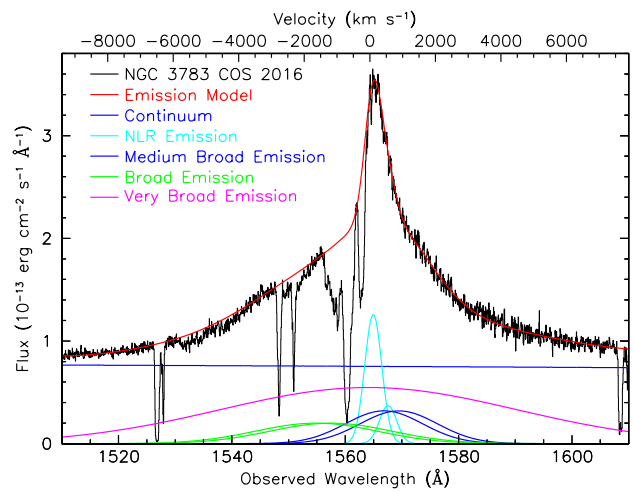}}
\caption{
Best fit with no absorption to the average COS spectrum of the \ion{C}{iv}
region from observations in 2016 December.
The data are the black histogram.
The solid red line tracing the data is the total emission model.
The key in the figure identifies the emission components in our model.
The velocity scale along the top axis is for the blue component of the
\ion{C}{iv} doublet, $\lambda1548.195$, relative to the host galaxy systemic
redshift, z=0.00973 \citep{Theureau98}.
}
\label{fig_cos_c4alt}
\end{figure}

Given that there are changes in the \ion{C}{iv} profile that do not strictly
mimic our comparison of the \ion{He}{ii} profiles, one might legitimately ask
whether the COS 2016 \ion{C}{iv} profile could be modeled without invoking
absorption components, but allowing for more dramatic changes in the
emission-line structure. To test this possibility, we made interactive
adjustments to the line centers, widths, and fluxes of the narrow,
medium-broad, broad, and very broad components in the \ion{C}{iv} profile.
We also included the portions of the line profile in the 1530--1545 \AA\ and
1552--1555 \AA\ regions which we had been treating as absorbed and required
them to be part of the emission profile.
We then let the fit try to optimize itself to the observed profile
without including any absorption.
Figure \ref{fig_cos_c4alt} shows the resulting fit.

This alternative interpretation of the line profile is not a good fit, and
it exhibits several unsatisfactory features.
First, there are smaller scale, shallow features on the blue side of the line
profile in the $-$2000 to $-$6000 $\rm km~s^{-1}$ velocity
range that are not fit well; these regions still look like shallow absorption
features. Second, the narrow emission components have to move in velocity to
+100 $\rm km~s^{-1}$, and they have an unphysical ratio of 3.4:1
for the blue to red intensity ratio, which should be $\leq$  2:1.
Finally, the medium-broad components have moved in velocity from near systemic
in the STIS spectrum to +550 $\rm km~s^{-1}$ here,
and the broad components have blue-shifted even more to $-$1650 $\rm km~s^{-1}$.

As we noted earlier, the \ion{He}{ii} profile does not show such dramatic
changes. We have tried to correct some of these shortcomings by fixing the
narrow emission components in flux and velocity at the STIS 2001 values,
forcing the medium-broad components to have zero systemic velocity
(as they did in the STIS spectrum), and then let everything else vary freely. 
As expected, the fit is even worse, and the irregular features on the blue
wing of the line profile simply can't be accommodated by such a model.
We conclude that absorption is the best explanation for the changes in the
shape of the \ion{C}{iv} profile.

\subsubsection{Modeling the Obscured Ly$\alpha$ Profile}

\begin{figure}[!tbp]
\centering
\resizebox{1.0\hsize}{!}{\includegraphics[angle=270]{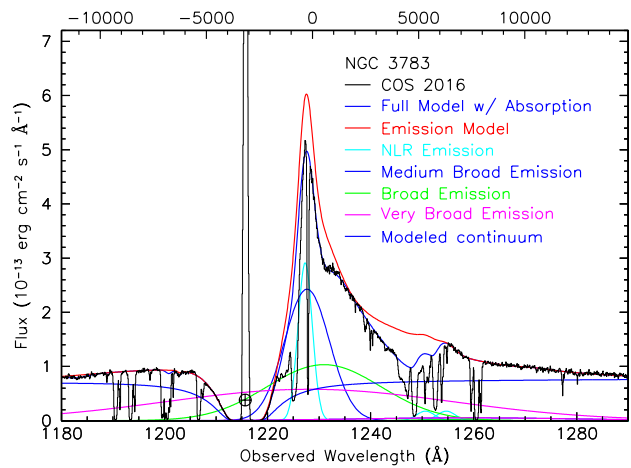}}
\caption{
Average COS spectrum of the Ly$\alpha$+\ion{N}{v} region from observations
in 2016 December (black histogram).
The solid red line tracing the data is the total emission model.
The solid blue line tracing the data is the emission model modified by
components that model the broad absorption.
The key in the figure identifies the emission components in our model.
Geocoronal emission in the center of the Milky Way
Ly$\alpha$ absorption trough is indicated with an Earth symbol.
The velocity scale along the top axis is for Ly$\alpha$
$\lambda 1215.67$, relative to the host galaxy systemic
redshift, z=0.00973 \citep{Theureau98}.
}
\label{fig_cos_lyan5all}
\end{figure}

To model the Ly$\alpha$ and \ion{N}{v} region of the COS spectrum, we start
with the emission model developed for the STIS 2001 spectrum.
As with \ion{C}{iv}, we keep the narrow emission components fixed in flux,
velocity, and width. We scale the broader components down in flux
to accommodate the decrease in overall flux.
Given the convincing evidence for broad, blue-shifted absorption affecting the
\ion{C}{iv} emission-line profile during the obscured state of NGC 3783, we
then impose a scaled replica of the modeled \ion{C}{iv} absorption to the
the Ly$\alpha$ and \ion{N}{v} regions. This initial model for the absorption
has the same velocities, widths, and optical depths as for \ion{C}{iv},
and provides a very good approximation to the new features present on the
red wing of Ly$\alpha$ (that are due to \ion{N}{v} absorption) as well as
the more subtle changes in the blue wing of Ly$\alpha$.
After selecting initial values by eye that give an approximate fit, we then
start freeing parameters and iterating to an overall solution as
described at the beginning of \S3.
For our final fit, all parameters vary freely.
Figure \ref{fig_cos_lyan5all} shows the resulting best fit to the
Ly$\alpha$ and \ion{N}{v} region.
The final model for the Ly$\alpha$ absorption profile resembles the
\ion{C}{iv} profile closely in velocity and width (see \S3.3),
but it is shallower in optical depth.

\subsubsection{Modeling the Obscured Si IV Profile}

\begin{figure}[!tbp]
\centering
\resizebox{1.0\hsize}{!}{\includegraphics[angle=270]{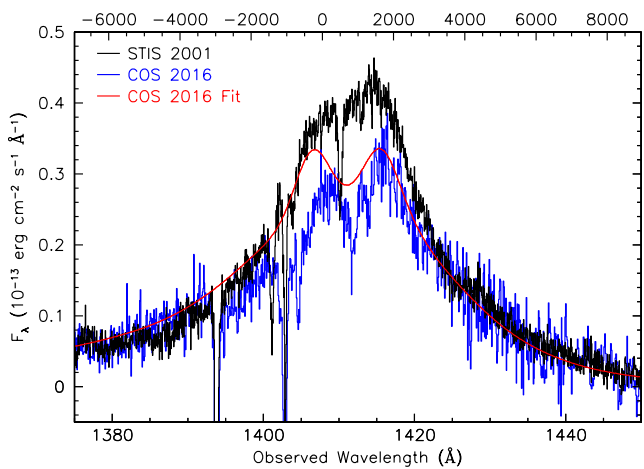}}
\caption{
Comparison of STIS and COS spectra surrounding the region of the \ion{Si}{iv}
emission line.
The black histogram is the average STIS 2001 spectrum.
The blue histogram is the unweighted average COS spectrum from 2016.
The solid red line is the best-fit emission model for the COS spectrum. 
The velocity scale along the top axis is for
\ion{Si}{iv} $\lambda1393.755$, relative to the host galaxy systemic
redshift, z=0.00973 \citep{Theureau98}.
}
\label{fig_stiscossi4all}
\end{figure}

Modeling the \ion{Si}{iv} region is important.
Since it is the lowest-ionization species to show significant absorption,
it sets a lower bound on the ionization state of the broad absorber.
Unlike in NGC 5548 \citep{Arav15}, absorption from lower ionization species
such as \ion{Si}{ii} or \ion{C}{ii} are not present in our spectra.
Fits to the \ion{Si}{iv} region, however, are more ambiguous since the line is
not very bright,
and the S/N is not as good as in the \ion{C}{iv} and Ly$\alpha$ regions.
\ion{Si}{iv} is also blended with several \ion{O}{iv}] transitions;
however, these appear to be weak, and as in the STIS spectrum, they primarily
affect rest wavelengths $> 1403$ \AA.
Starting again with the model for the STIS 2001 spectrum,
Figure \ref{fig_stiscossi4all} compares the STIS and COS
continuum-subtracted line profiles.
No scaling has been applied to either spectrum. Again, we have the remarkable
result that despite the continuum being twice as bright in 2016 as in 2001,
the emission line in 2001 is slightly brighter.
As with \ion{C}{iv}, most of this brighter emission is in the core of the line.

\begin{figure}[!tbp]
\centering
\resizebox{1.0\hsize}{!}{\includegraphics[angle=270]{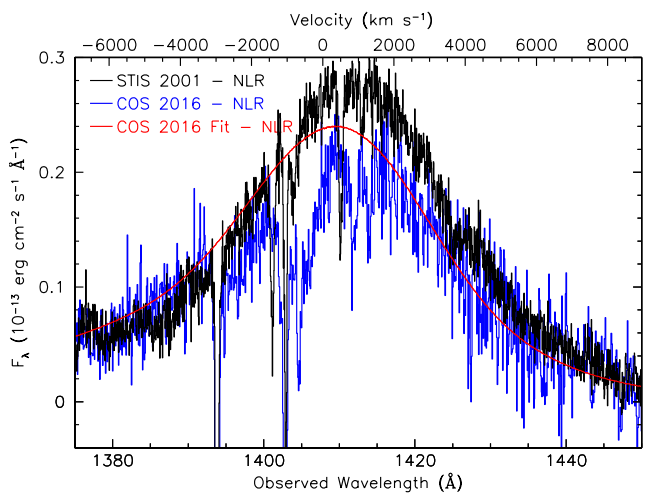}}
\caption{
Comparison of STIS and COS spectra of the region surrounding the \ion{Si}{iv}
emission line omitting the narrow emission components.
The black histogram is the average STIS 2001 spectrum with the narrow emission
components of \ion{Si}{iv} and \ion{O}{iv}] subtracted.
The blue histogram is the unweighted average COS spectrum from 2016, also
with the narrow emission components subtracted.
The solid red line is the best-fit emission model for the COS spectrum
minus the narrow emission components. 
The velocity scale along the top axis is for
\ion{Si}{iv} $\lambda1393.755$, relative to the host galaxy systemic
redshift, z=0.00973 \citep{Theureau98}.
}
\label{fig_stiscossi4n}
\end{figure}

As for \ion{C}{iv}, we next subtract the narrow emission components of
\ion{Si}{iv} and \ion{O}{iv}].
Figure \ref{fig_stiscossi4n} shows this net spectrum for both epochs.
A slight excess still remains near the line center.
Recall that we use only 3 Gaussians to model \ion{Si}{iv} since it is so weak.
This remaining excess present in the STIS spectrum is analogous to what we have
modeled as medium-broad emission in the \ion{C}{iv} profile.
Removing this would then bring the STIS spectrum almost into line with the
model fit to the broad plus very broad emission in the COS 2016 spectrum. 

\begin{figure}[!tbp]
\centering
\resizebox{1.0\hsize}{!}{\includegraphics[angle=270]{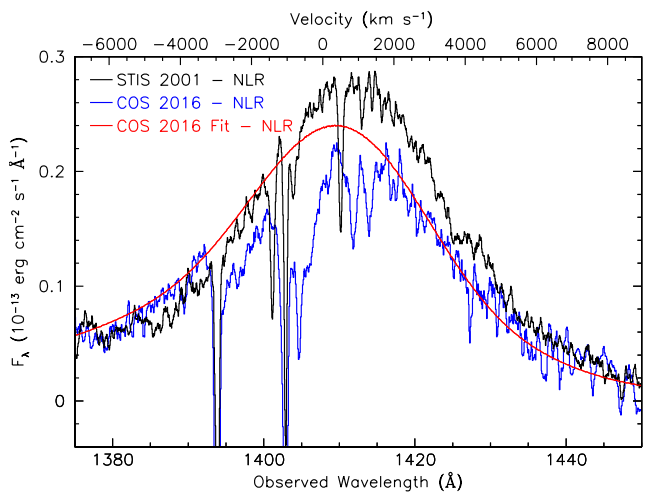}}
\caption{
Spectra of the \ion{Si}{iv} region as in Figure \ref{fig_stiscossi4n},
but with both the STIS and the COS spectra
smoothed by a 7-pixel running boxcar filter.
The solid red line is the best-fit emission model for the COS spectrum
minus the narrow emission components. 
The velocity scale along the top axis is for
\ion{Si}{iv} $\lambda1393.755$, relative to the host galaxy systemic
redshift, z=0.00973 \citep{Theureau98}.
}
\label{fig_stiscossi4ns}
\end{figure}

Is there broad absorption present in the COS 2016 spectrum?
Near the line peak this is uncertain since the narrow absorption lines are
deeper and stronger than in 2001, and they may be responsible for most of the
diminution in flux we observe.
However, at velocities in the blue wing from $-$1500 $\rm km~s^{-1}$ to $-$3000
$\rm km~s^{-1}$,
there is an obvious depression in flux below the model profile.
Smoothing both spectra with a 7-pixel running boxcar filter as shown in
Figure \ref{fig_stiscossi4ns} shows this depression more clearly.
The model, however, is far from unique.
Given the S/N, an acceptable fit is possible where the only absorption is
due to enhanced absorption in the narrow absorption lines.
The broad absorption shown in Figures \ref{fig_stiscossi4n} and
\ref{fig_stiscossi4ns} should therefore be considered upper limits.
Relative to the continuum, the depression in flux at $-$2500 $\rm km~s^{-1}$
corresponds to an optical depth of only 0.05.

\begin{sidewaystable*}
\begin{center}
        \caption[]{Emission-Line Parameters for the Individual COS 2016 Spectra of NGC~3783}
        \label{tab:cos3cos4_em}
\begin{tabular}{l c c c c c c c c}
\hline\hline
        &      & \multicolumn{3}{c}{COS 2016-12-12} &   &  \multicolumn{3}{c}{COS 2016-12-21}   \\
Feature & $\rm \lambda_0^{\mathrm{a}}$ &  $\rm Flux^{\mathrm{b}}$ & $\rm v_{sys}^{\mathrm{c}}$ & $\rm FWHM^{\mathrm{d}}$ & \phantom{000} & $\rm Flux^{\mathrm{b}}$ & $\rm v_{sys}^{\mathrm{c}}$ & $\rm FWHM^{\mathrm{d}}$ \\
\hline
\ion{C}{iii}\*         &   1176.01 & $  47.0\pm  2.6$ & $   170 \pm 110$ & $ 1490 \pm   50$ & \phantom{000} & $  34.0\pm  1.5$ & $   130 \pm  20$ & $ 1520 \pm   60$\\
Ly$\alpha$             &   1215.67 & $ 120.0\pm  4.4$ & $   -40 \pm  30$ & $  800 \pm   30$ & \phantom{000} & $ 120.0\pm  4.0$ & $   -40 \pm  20$ & $  800 \pm   20$\\
Ly$\alpha$             &   1215.67 & $ 250.0\pm  8.2$ & $  -240 \pm  40$ & $ 4190 \pm   40$ & \phantom{000} & $ 340.0\pm 12.0$ & $  -240 \pm  20$ & $ 4190 \pm   40$\\
Ly$\alpha$             &   1215.67 & $ 350.0\pm 11.0$ & $   -70 \pm  30$ & $ 6950 \pm   20$ & \phantom{000} & $ 310.0\pm 13.0$ & $   -70 \pm  30$ & $ 6950 \pm   70$\\
Ly$\alpha$             &   1215.67 & $ 320.0\pm  9.8$ & $    60 \pm  40$ & $18010 \pm   70$ & \phantom{000} & $ 320.0\pm 11.0$ & $    60 \pm  20$ & $18010 \pm   30$\\
\ion{N}{v} blue        &   1238.82 & $   5.1\pm  0.2$ & $   -40 \pm  30$ & $  980 \pm   30$ & \phantom{000} & $   7.8\pm  1.2$ & $   -40 \pm  30$ & $  980 \pm   50$\\
\ion{N}{v} red         &   1242.80 & $   5.1\pm  0.2$ & $   -40 \pm  30$ & $  980 \pm   30$ & \phantom{000} & $   7.8\pm  1.2$ & $   -40 \pm  30$ & $  980 \pm   50$\\
\ion{N}{v} blue        &   1238.82 & $   7.4\pm  0.3$ & $   -30 \pm  30$ & $ 2850 \pm   40$ & \phantom{000} & $   9.9\pm  0.6$ & $   -30 \pm  60$ & $ 2850 \pm   20$\\
\ion{N}{v} red         &   1242.80 & $   7.4\pm  0.3$ & $   -30 \pm  30$ & $ 2850 \pm   40$ & \phantom{000} & $   9.9\pm  0.6$ & $   -30 \pm  60$ & $ 2850 \pm   20$\\
\ion{N}{v}             &   1240.89 & $  25.0\pm  1.0$ & $  -220 \pm 170$ & $12530 \pm  530$ & \phantom{000} & $  73.0\pm  2.6$ & $  -220 \pm  50$ & $12530 \pm   40$\\
\ion{Si}{ii}            &   1260.42 & $   1.5\pm  0.1$ & $  -150 \pm  30$ & $ 1600 \pm   40$ & \phantom{000} & $   1.7\pm  0.2$ & $  -150 \pm  50$ & $ 1600 \pm  120$\\
\ion{O}{i}+\ion{Si}{ii} &   1304.46 & $  10.0\pm  1.0$ & $   220 \pm  30$ & $ 2500 \pm   50$ & \phantom{000} & $  11.0\pm  0.4$ & $   220 \pm  40$ & $ 2500 \pm   60$\\
\ion{C}{ii}            &   1334.53 & $   2.7\pm  0.2$ & $   110 \pm  40$ & $ 2500 \pm   20$ & \phantom{000} & $   2.8\pm  0.2$ & $   110 \pm  50$ & $ 2500 \pm   20$\\
\ion{Si}{iv} blue       &   1393.76 & $   4.6\pm  0.2$ & $  -170 \pm  50$ & $ 1110 \pm   40$ & \phantom{000} & $   4.6\pm  0.2$ & $  -170 \pm  30$ & $ 1110 \pm   70$\\
\ion{Si}{iv} red        &   1402.77 & $   4.6\pm  0.2$ & $  -170 \pm  50$ & $ 1110 \pm   20$ & \phantom{000} & $   4.6\pm  0.2$ & $  -170 \pm  30$ & $ 1110 \pm   70$\\
\ion{Si}{iv} blue       &   1393.76 & $  17.0\pm  0.8$ & $  -270 \pm  20$ & $ 5140 \pm   60$ & \phantom{000} & $  28.0\pm  1.2$ & $  -270 \pm  30$ & $ 5140 \pm   80$\\
\ion{Si}{iv} red        &   1402.77 & $  17.0\pm  0.8$ & $  -270 \pm  20$ & $ 5140 \pm   60$ & \phantom{000} & $  28.0\pm  1.2$ & $  -270 \pm  30$ & $ 5140 \pm   80$\\
\ion{Si}{iv}            &   1398.19 & $  62.0\pm  2.3$ & $ -2750 \pm  30$ & $13030 \pm   80$ & \phantom{000} & $  51.0\pm  1.7$ & $ -2750 \pm  40$ & $13030 \pm  140$\\
\ion{O}{iv}]           &   1401.16 & $   0.7\pm  0.1$ & $   900 \pm  30$ & $ 1110 \pm   20$ & \phantom{000} & $   0.7\pm  0.1$ & $   900 \pm  40$ & $ 1110 \pm   20$\\
\ion{O}{iv}]           &   1401.16 & $   9.1\pm  1.0$ & $   900 \pm  30$ & $ 5140 \pm   20$ & \phantom{000} & $   9.1\pm  0.4$ & $   900 \pm  40$ & $ 5140 \pm   20$\\
\ion{N}{iv}]           &   1486.50 & $   4.1\pm  0.4$ & $  -140 \pm  20$ & $ 1110 \pm   20$ & \phantom{000} & $   2.5\pm  0.3$ & $   -80 \pm  20$ & $ 1110 \pm   20$\\
\ion{N}{iv}]           &   1486.50 & $  11.0\pm  0.8$ & $  -140 \pm  20$ & $ 4330 \pm   60$ & \phantom{000} & $   6.3\pm  0.9$ & $   -80 \pm  20$ & $ 2690 \pm   20$\\
\ion{C}{iv} blue       &   1548.19 & $  35.0\pm  1.1$ & $   -60 \pm  20$ & $  940 \pm   20$ & \phantom{000} & $  38.0\pm  1.9$ & $   -60 \pm  20$ & $  940 \pm   30$\\
\ion{C}{iv} red        &   1550.77 & $  35.0\pm  1.1$ & $   -60 \pm  20$ & $  940 \pm   20$ & \phantom{000} & $  38.0\pm  1.9$ & $   -60 \pm  20$ & $  940 \pm   30$\\
\ion{C}{iv} blue       &   1548.19 & $   3.3\pm  0.2$ & $   140 \pm  50$ & $ 2840 \pm   20$ & \phantom{000} & $   4.8\pm  0.3$ & $   140 \pm  20$ & $ 2840 \pm   70$\\
\ion{C}{iv} red        &   1550.77 & $   3.3\pm  0.2$ & $   140 \pm  50$ & $ 2840 \pm   20$ & \phantom{000} & $   4.8\pm  0.3$ & $   140 \pm  20$ & $ 2840 \pm   70$\\
\ion{C}{iv} blue       &   1548.19 & $ 190.0\pm  5.7$ & $  -790 \pm  20$ & $ 4800 \pm   20$ & \phantom{000} & $ 160.0\pm  5.0$ & $  -570 \pm  30$ & $ 4530 \pm   30$\\
\ion{C}{iv} red        &   1550.77 & $ 190.0\pm  5.7$ & $  -790 \pm  20$ & $ 4800 \pm   20$ & \phantom{000} & $ 160.0\pm  5.0$ & $  -570 \pm  30$ & $ 4530 \pm   30$\\
\ion{C}{iv}            &   1549.48 & $ 310.0\pm 11.0$ & $  -220 \pm  30$ & $13000 \pm  120$ & \phantom{000} & $ 370.0\pm 13.0$ & $   120 \pm  30$ & $12500 \pm   50$\\
\ion{He}{ii}           &   1640.45 & $  14.0\pm  0.7$ & $    40 \pm  20$ & $  910 \pm   30$ & \phantom{000} & $  14.0\pm  1.4$ & $     0 \pm  30$ & $  790 \pm   30$\\
\ion{He}{ii}           &   1640.45 & $   7.9\pm  0.3$ & $  2100 \pm 100$ & $ 5090 \pm  120$ & \phantom{000} & $   8.7\pm  0.4$ & $  1400 \pm 110$ & $ 3490 \pm  270$\\
\ion{He}{ii}           &   1640.45 & $ 150.0\pm  4.8$ & $  -120 \pm  30$ & $11860 \pm   60$ & \phantom{000} & $ 170.0\pm  5.2$ & $  -120 \pm  20$ & $12530 \pm  100$\\
\ion{O}{iii}]          &   1660.81 & $   2.3\pm  0.3$ & $   160 \pm  50$ & $ 1220 \pm   60$ & \phantom{000} & $   3.2\pm  0.1$ & $   160 \pm  30$ & $ 1210 \pm   30$\\
\ion{O}{iii}]          &   1666.15 & $   4.3\pm  0.5$ & $   150 \pm  20$ & $ 1220 \pm   20$ & \phantom{000} & $   6.8\pm  0.4$ & $   160 \pm  20$ & $ 1210 \pm   20$\\
\ion{N}{iii}]          &   1750.00 & $  21.0\pm  0.8$ & $     0 \pm  20$ & $ 3270 \pm   30$ & \phantom{000} & $  21.0\pm  0.6$ & $     0 \pm  20$ & $ 3270 \pm   20$\\
\hline
\end{tabular}
\end{center}
{\bf Notes.}\\
$^{\mathrm{a}}$ Vacuum rest wavelength of the spectral feature (\AA).\\
$^{\mathrm{b}}$ Integrated flux in units of $\rm 10^{-14}~erg~cm^{-2}~s^{-1}$.\\
$^{\mathrm{c}}$ Velocity (in $\rm km~s^{-1}$) relative to a systemic redshift of z = 0.00973 \citep{Theureau98}.\\
$^{\mathrm{d}}$ Full-width at half-maximum ($\rm km~s^{-1}$).\\
\end{sidewaystable*}

\subsubsection{Modeling the Obscured H$\beta$ Profile}

\begin{figure}[!tbp]
\centering
\resizebox{1.0\hsize}{!}{\includegraphics[angle=270]{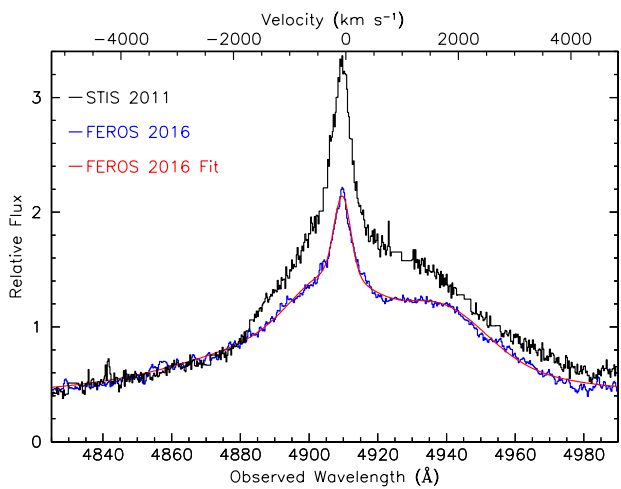}}
\caption{
FEROS spectrum of the H$\beta$ region from observations in 2016
December (blue histogram) compared to the STIS spectrum of 2011 (black).
The continuum has been subtracted from each spectrum, and the FEROS spectrum
has been scaled to match the flux in the far blue wing of the STIS spectrum.
The solid red line tracing the FEROS data is the total emission model with
the continuum subtracted.
The velocity scale along the top axis is for the rest wavelength of H$\beta$
relative to the host galaxy systemic
redshift, z=0.00973 \citep{Theureau98}.
}
\label{fig_feros_hball}
\end{figure}

Like \ion{He}{ii}, H$\beta$ emission originates as recombination radiation to
an excited level. Similarly, we do not expect absorption in such a feature.
As our baseline for comparison, we use a STIS spectrum obtained in 2011 when
NGC 3783 was in an unobscured state.
Our FEROS spectrum was obtained on 2016 December 12,
simultaneously with our first {\it XMM-Newton} spectrum of NGC 3783 in its
obscured state. 
Analogous to our fit to the \ion{C}{iv} region, we fit both the STIS and the
FEROS H$\beta$ spectra using a power law for the continuum,
a narrow emission component, a medium-broad component, a broad component, and
a very broad component.
Significantly, both spectra also require an additional emission bump on the
red side of the line profile, although the bump is more prominent in the
FEROS spectrum.
Figure \ref{fig_feros_hball} compares the unobscured STIS spectrum of the
H$\beta$ region to the FEROS spectrum.
As in our previous comparisons, we have subtracted the continuum and scaled
the FEROS spectrum to match the flux level in the far blue wing of the
emission-line profile.

\begin{figure}[!tbp]
\centering
\resizebox{1.0\hsize}{!}{\includegraphics[angle=270]{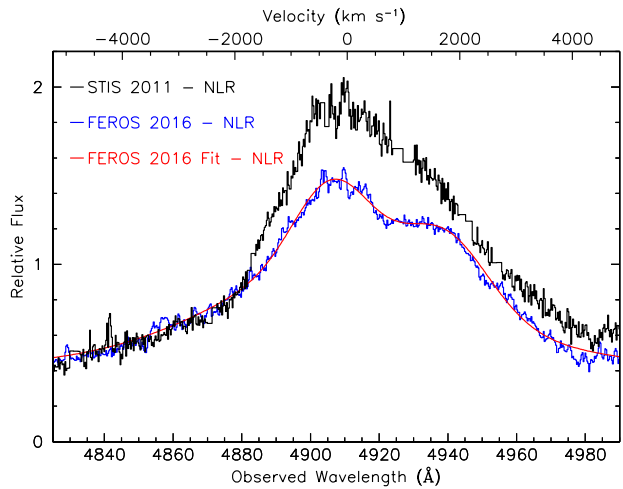}}
\caption{
FEROS (blue) and STIS (black) spectra of the H$\beta$ region with the continuum
and narrow H$\beta$ emission component subtracted from each.
The FEROS spectrum has been scaled to match the STIS spectrum in the far blue
wing of the emission line.
The solid red line tracing the data is the total emission model with the
continuum and narrow emission component subtracted.
The velocity scale along the top axis is for the rest wavelength of H$\beta$
relative to the host galaxy systemic
redshift, z=0.00973 \citep{Theureau98}.
}
\label{fig_feros_hbnarrow}
\end{figure}

\begin{figure}[!tbp]
\centering
\resizebox{1.0\hsize}{!}{\includegraphics[angle=270]{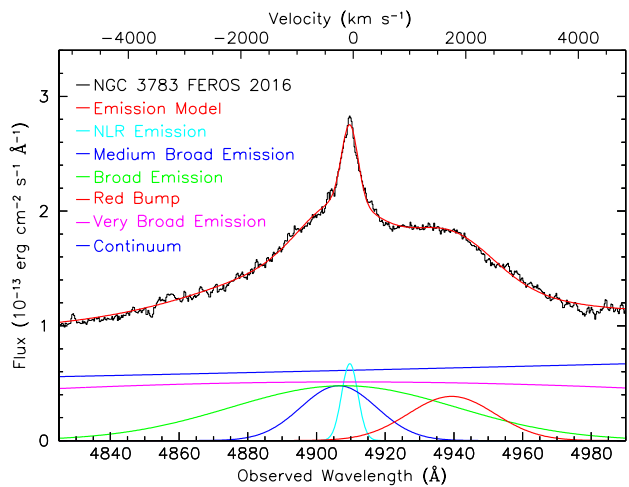}}
\caption{
FEROS spectrum of the H$\beta$ region from observations in 2016
December (black histogram).
The solid red line tracing the data is the total emission model.
The key in the figure identifies the emission components in our model.
The velocity scale along the top axis is for the rest wavelength of H$\beta$
relative to the host galaxy systemic
redshift, z=0.00973 \citep{Theureau98}.
}
\label{fig_feros_hbfit}
\end{figure}

In Figure \ref{fig_feros_hbnarrow}, we compare the profiles with both the
continuum and the narrow emission component subtracted.
The two profiles now show behavior similar to our previous emission-line
comparisons--the core of the emission line in the unobscured state is brighter.
In the obscured state, this core seems to have faded, analogous to the
disappearance of the medium-broad component in \ion{C}{iv}. This makes the
red emission bump in the H$\beta$ line profile more prominent in the
obscured state. These changes can be accommodated simply by changing the
relative fluxes of the emission components. There is no need for any
absorption in the line profile.
All the components in the final best-fit to the FEROS H$\beta$ emission line
profile are illustrated in Figure \ref{fig_feros_hbfit}.

\subsection{Photoionization Modeling of the Broad Absorption Lines}

\begin{figure}[!tbp]
\vspace{-0.25cm}
\centering
\hspace*{-0.25cm}\resizebox{1.05\hsize}{!}{\includegraphics[angle=0]{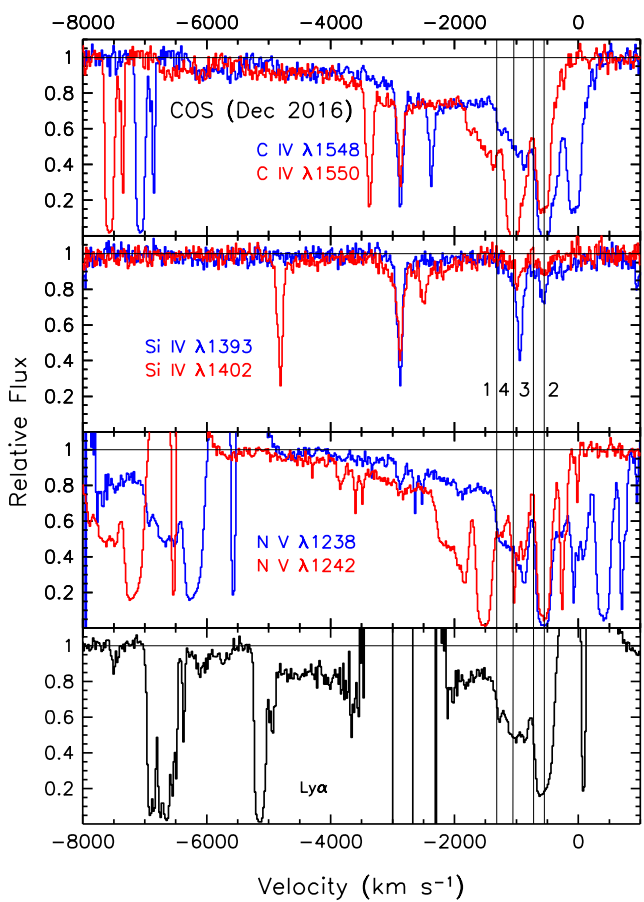}}
\caption{Normalized spectra of the broad absorption features in NGC~3783
(2016-12-12).
The velocities of components 1--4 as given by \cite{Gabel03b} are marked by
thin vertical blue lines and labeled.
}
\label{fig_broadnorm03}
\end{figure}

Our models of the emission and broad absorption in NGC 3783 now allow us to
measure the physical properties of the gas causing the broad absorption and
possibly also the soft X-ray obscuration observed in 2016 December.
Using our model for the emission spectrum, we produce normalized spectra for
the regions affected by the broad UV absorption, as shown in
Figure \ref{fig_broadnorm03}.

We can integrate our normalized models for the broad absorption using the
apparent optical depth method of \cite{Savage91} to obtain column densities,
$\rm N_{ion}$.
However, since the lines appear to be saturated, these column densities are
only lower limits to the true column density.
In Table \ref{tab:broadabscolumns} we summarize the observable properties of
the modeled broad absorption troughs.
The equivalent width (EW) is integrated from the normalized spectra between
the velocity limits $v_1$ and $v_2$.
The deepest point of the trough is at velocity $v_o$, and we use the dispersion
$\sigma_v$ to characterize the width of the trough.
The covering fraction, $\rm C_f$, is measured by assuming that the profile is
saturated at the deepest point of the trough.

\begin{table*}
  \caption[]{Properties of the Broad Absorption Troughs in COS Observations of NGC~3783.
     $\lambda_o$ is the vacuum rest wavelength of the spectral feature. Absorption troughs
     span the velocity range given by {$v_1$} to {$v_2$}. {$v_o$} is the
     transmission-weighted velocity centroid of the trough, EW is its equivalent width,
     and $\rm log(N_{ion})$ is the inferred ionic column density.
     Assuming the trough is saturated, $\rm C_f$ is the covering factor at the
     deepest point in the absorption trough.
  }
  \label{tab:broadabscolumns}
\begin{center}
\begin{tabular}{l c c c c c c c c}
\hline\hline
{Line} & $\lambda_o$ & {$v_1$} & {$v_2$} & {$v_o$} & {$\sigma_v$} & {EW} & $\rm log(N_{ion})$ & $\rm C_f$ \\
      & (\AA)       &  ($\rm km~s^{-1}$) & ($\rm km~s^{-1}$) &  ($\rm km~s^{-1}$) &  ($\rm km~s^{-1}$) & (\AA) & ($\rm cm^{-2}$) & \\
\hline
{COS 2016-12-12} \\
\hline
Ly$\alpha$   & 1215.67  & $-6960$ & 0 & $-3000$ & $\phantom{0}830$ & $2.16 \pm 0.08$ & $>14.58$ & 0.18 \\
\ion{N}{v}   & 1240.51  & $-6600$ & 0 & $-1300$ & $1050$ & $3.10 \pm 0.11$ & $>15.66$ & 0.37 \\
\ion{Si}{ii} & 1260.42  & $-6810$ & 0 & --- & --- & $<0.22$ & $<13.63$ & --- \\
\ion{C}{ii}  & 1334.53  & $-6810$ & 0 & --- & --- & $<0.18$ & $<14.22$ & --- \\
\ion{Si}{iv} & 1398.27  & $-7620$ & 0 & $-3000$ & $2300$ & $1.33 \pm 0.09$ & $>14.48$ & 0.08 \\
\ion{C}{iv}  & 1549.48  & $-6810$ & 0 & $-3000$ & $1470$ & $6.13 \pm 0.21$ & $>15.51$ & 0.30 \\
\hline
{COS 2016-12-21} \\
\hline
Ly$\alpha$   & 1215.67  & $-6960$ & 0 & $-3000$ & $\phantom{0}860$ & $1.25 \pm 0.05$ & $>14.35$ & 0.12 \\
\ion{N}{v}   & 1240.51  & $-6600$ & 0 & $-1300$ & $\phantom{0}880$ & $2.38 \pm 0.09$ & $>15.56$ & 0.33 \\
\ion{Si}{ii} & 1260.42  & $-6810$ & 0 & --- & --- & $<0.18$ & $<13.68$ & --- \\
\ion{C}{ii}  & 1334.53  & $-6810$ & 0 & --- & --- & $<0.20$ & $<14.13$ & --- \\
\ion{Si}{iv} & 1398.27  & $-7620$ & 0 & $-3000$ & $1680$ & $1.61 \pm 0.10$ & $>14.55$ & 0.11 \\
\ion{C}{iv}  & 1549.48  & $-6810$ & 0 & $-3000$ & $1320$ & $4.05 \pm 0.15$ & $>15.31$ & 0.23 \\
\hline
\end{tabular}
\end{center}
\end{table*}

Unlike the obscured state of NGC 5548 \citep{Kaastra14, Arav15}, where
absorption from low-ionization species such as
\ion{C}{ii} $\lambda 1335$ and \ion{Si}{iii} $\lambda1206$
was present, in the obscured state of NGC 3783 the lowest ionization
broad absorption feature is \ion{Si}{iv}.
To model the ionization state of the absorbing clouds, we use the unobscured
spectral energy distribution (SED) of NGC 3783 as presented in the top panel of
Figure 6 of \cite{Mehdipour17}.
We derive the ionization balance of the obscurer using a grid of models
generated using the Cloudy v17.00 photoionization code \citep{Ferland17}.
Our grid covers a range in ionization parameter log $\xi$ ($\rm erg~cm~s^{-1}$)
from 1.0 to 2.5,
and total column density $\rm log~N_H~(cm^{-2})$ from 21.5 to 23.7.
The ionization parameter has the usual definition, $\xi = L_{ion} / (n r^2)$,
where $L_{ion}$ is the ionizing luminosity from 1 to 1000 Ryd, $n$ is the
density, and $r$ is the distance of the absorbing cloud from the
ionization source.

\begin{figure}[!tbp]
\centering
\resizebox{1.0\hsize}{!}{\includegraphics[angle=0]{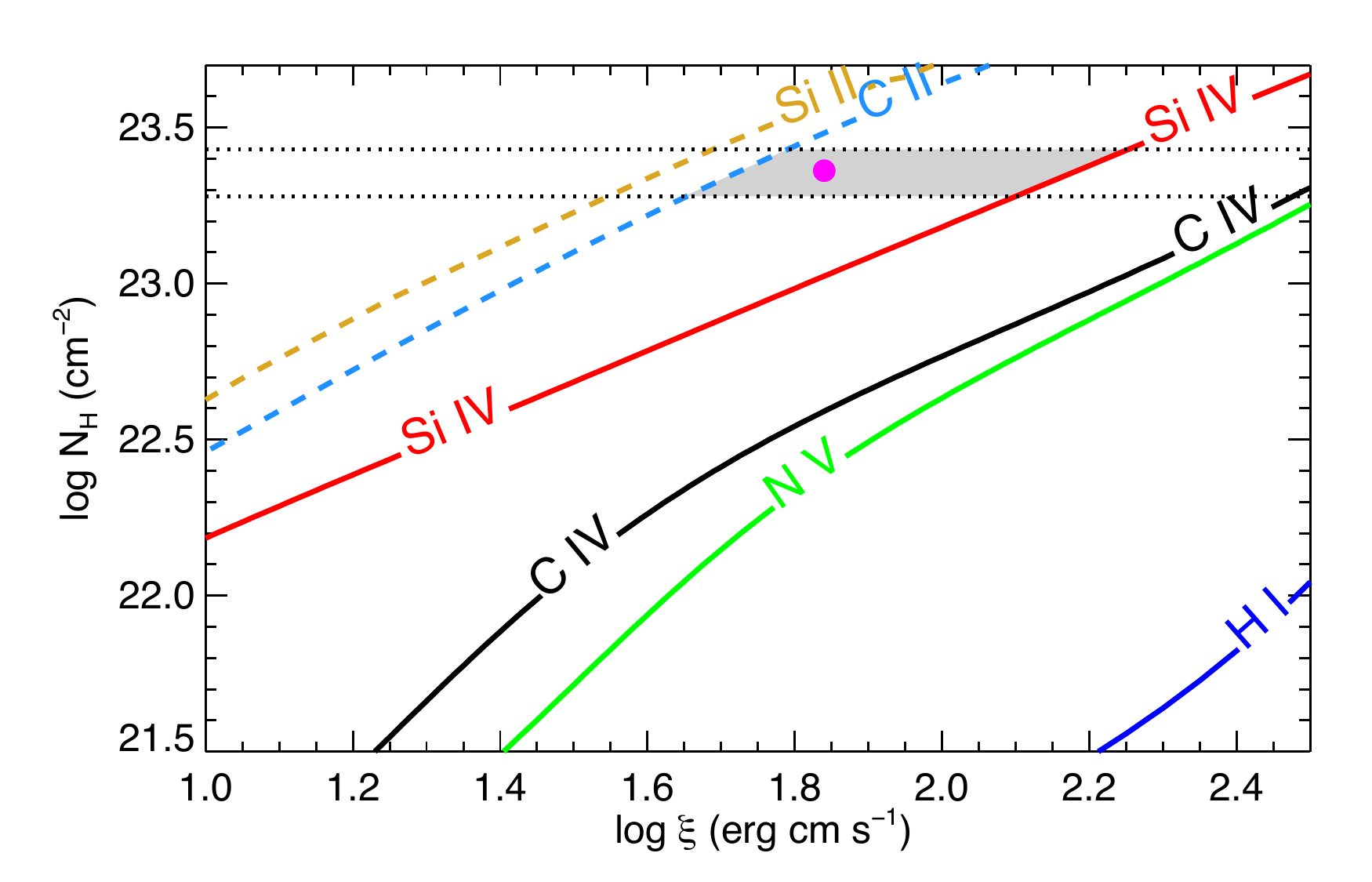}}
\caption{
Constraints on photoionization models for the obscurer in NGC~3783.
Dotted black lines give the constraints on total column density
allowed by our X-ray spectra in \cite{Mehdipour17}.
Solid colored lines specify lower limits on the column densities of the
indicated ions. Allowed photoionization solutions lie in the space
above these lines.
Dashed colored lines give upper limits on the column densities of the indicated
low-ionization ions. Allowed photoionization solutions lie in the space
below these lines.
The shaded parallelogram shows the allowed parameter space, and the
magenta dot gives the solution used for fitting the X-ray spectra in
\cite{Mehdipour17}.
}
\label{fig:nion_xi}
\end{figure}

For absorption troughs detected in our spectra, Ly$\alpha$, \ion{N}{v},
\ion{C}{iv}, and \ion{Si}{iv}, we use the lower limits on the
total column density given in Table \ref{tab:broadabscolumns}
and show these limits as solid lines in Figure \ref{fig:nion_xi}.
Using a transmission profile with the same shape in velocity as \ion{C}{iv},
we allow the optical depths at the locations of
\ion{C}{ii} $\lambda 1335$ and \ion{Si}{ii} $\lambda1260$ to vary until
$\chi^2$ increases by 4.0 above its minimum value. This then gives us
2$\sigma$ upper limits (for a single interesting parameter)
on their column densities, which are give in
Table \ref{tab:broadabscolumns}, and
shown as dashed lines in Figure \ref{fig:nion_xi}.
(We note that \ion{Si}{iii} $\lambda1206$ in NGC 3783 is buried in the damped Ly$\alpha$ Milky Way absorption.)
Acceptable photoionization solutions for the broad absorbing gas in
NGC~3783 should have ionization parameters and column densities in the region
above the solid lines and below the dashed lines.
A solid magenta dot in Figure \ref{fig:nion_xi} shows the photoionization
solution used by \cite{Mehdipour17} for fitting the X-ray spectrum of
NGC~3783. Due to the lack of associated X-ray spectral features, this
solution is unconstrained in ionization parameter, but it is tightly
constrained in total column density, as shown by the bracketing dashed
black lines. Thus the combined UV and X-ray data constrain the
photoionization state of the obscuring gas to the small, approximately
quadrilateral region surrounding the magenta dot in the figure.

\subsection{Variability of the Narrow Absorption Lines}\label{narrow_line_var}

The intrinsic narrow absorption lines in NGC 3783 are prominent features
in its UV spectrum. The proximity and brightness of NGC 3783 has made it
a favorite target for trying to understand the physical characteristics
and origin of such intrinsic UV absorption features, and their
relationship to the blue-shifted X-ray absorption lines comprising the
X-ray warm absorber.
The extensive {\it Chandra}, {\it HST}, and {\it FUSE} monitoring
campaign in 2000--2001 \citep{Kaspi02, Gabel03} established the
baseline characteristics of these absorbers.
There are four discrete velocity components in the UV.
We designate them as \#1 through \#4 using the nomenclature established by
\cite{Gabel03}. Adjusting the \cite{Gabel03} velocities to the zero point
of the more precise redshift of \cite{Theureau98}, we will refer to the
four components as \#1 ($-1311~\rm km~s^{-1}$),
\#2 ($-539~\rm km~s^{-1}$),
\#3 ($-715~\rm km~s^{-1}$),
and \#4 ($-1018~\rm km~s^{-1}$).
\cite{Gabel05b} detected variations in
all four components in response to continuum variations as expected for
photoionized gas. These variations gave upper limits on the distance of the
absorbers in the range 25--50 pc; Component \#1 is located more precisely
since the density of log $\rm n_e = 4.5~cm^{-3}$ determined using
metastable \ion{C}{iii}* $\lambda1176$ places the absorbing gas at 25 pc.
The 15-year baseline of high-resolution {\it HST} spectra of NGC 3783
enables us to examine the behavior of the intrinsic narrow absorption
lines in more detail, and especially in the context of how they have
been affected by the obscuration event in 2016.

\subsubsection{Kinematics}

The variability of the narrow absorption lines in NGC 3783 has been both a
boon for our understanding of the UV outflow as well as an enigma.
In the extensive 2000--2001 monitoring campaign with STIS,
the low-ionization portion of Component \#1, designated \#1a by \cite{Gabel05b},
showed both absorption in the density-sensitive metastable transitions of
\ion{C}{iii}* $\lambda 1176$ as well as flux-dependent variability in
\ion{Si}{iv} that enabled a reliable measurement of the density of the
absorbing gas.
The puzzling aspect, however, is that Component \#1 appeared to change in
velocity in the sense that it decelerated over the course of the
2000--2001 campaign \citep{Gabel03b}.
Component \#1 appeared to move redward by $90~\rm km~s^{-1}$, from
$-1352~\rm km~s^{-1}$ to $-1256~\rm km~s^{-1}$.
\cite{Scott14} showed that this motion appeared to continue in their 2013
spectrum, with Component \#1 moving near to the apparent location of
Component \#4 at $-1100~\rm km~s^{-1}$.
They note that there is likely motion in other components as well.
Component \#3 has disappeared from its original location, and seems to have
moved redward into the region originally occupied by Component \#2, at least
to the blue side of the original trough.  However, there is no indication
that Component \#2 itself has moved redward.

\begin{figure}[!tbp]
\centering
\resizebox{1.0\hsize}{!}{\includegraphics[angle=270]{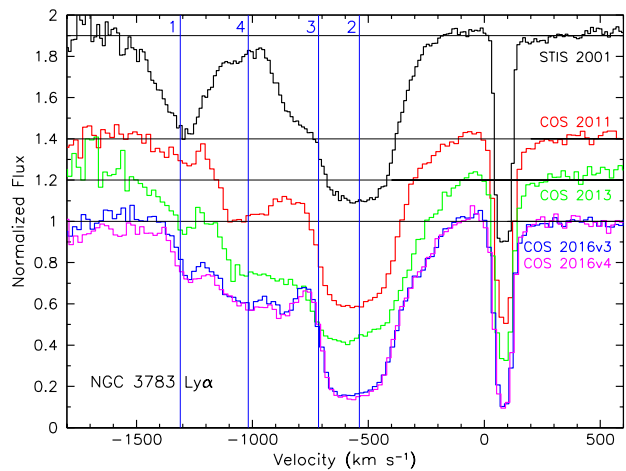}}
\caption{
Normalized HST spectra of the Ly$\alpha$ region from five different epochs
as labeled.
Velocity is for Ly$\alpha$ $\lambda$1215.67 relative to the host
galaxy systemic redshift of $z = 0.00973$ \citep{Theureau98}.
Fluxes are normalized to range from 0 to 1. Epochs prior to 2016 are offset
vertically by 0.2, 0.4, and 0.9, with the offset normalizations indicated
by thin horizontal black lines.
The velocities of components 1--4 as given by \cite{Gabel03b} are marked by
thin vertical blue lines and labeled.
}
\label{fig_lya_var}
\end{figure}

\begin{figure}[!tbp]
\centering
\resizebox{1.0\hsize}{!}{\includegraphics[angle=270]{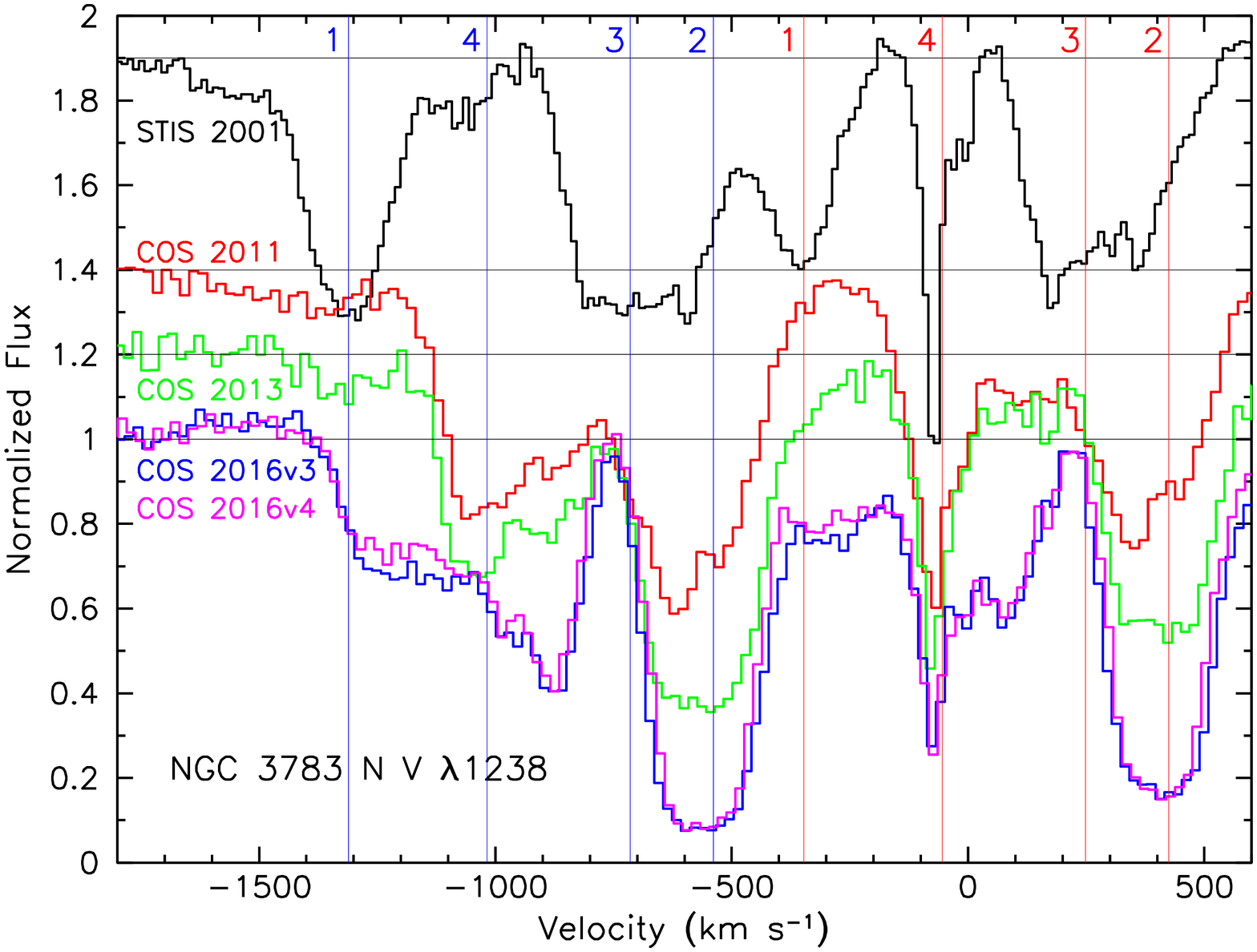}}
\caption{
Normalized HST spectra of the \ion{N}{v} region from five different epochs
as labeled.
Velocity is for \ion{N}{v} $\lambda$1238.821 relative to the host
galaxy systemic redshift of $z = 0.00973$ \citep{Theureau98}.
Fluxes are normalized to range from 0 to 1. Epochs prior to 2016 are offset
vertically by 0.2, 0.4, and 0.9, with the offset normalizations indicated
by thin horizontal black lines.
The velocities of components 1--4 as given by \cite{Gabel03b} are marked by
thin vertical blue lines and labeled; thin vertical red lines mark the expected
locations of the red component of the \ion{N}{v} doublet.
}
\label{fig_n5_var}
4\end{figure}

\begin{figure}[!tbp]
\centering
\resizebox{1.0\hsize}{!}{\includegraphics[angle=270]{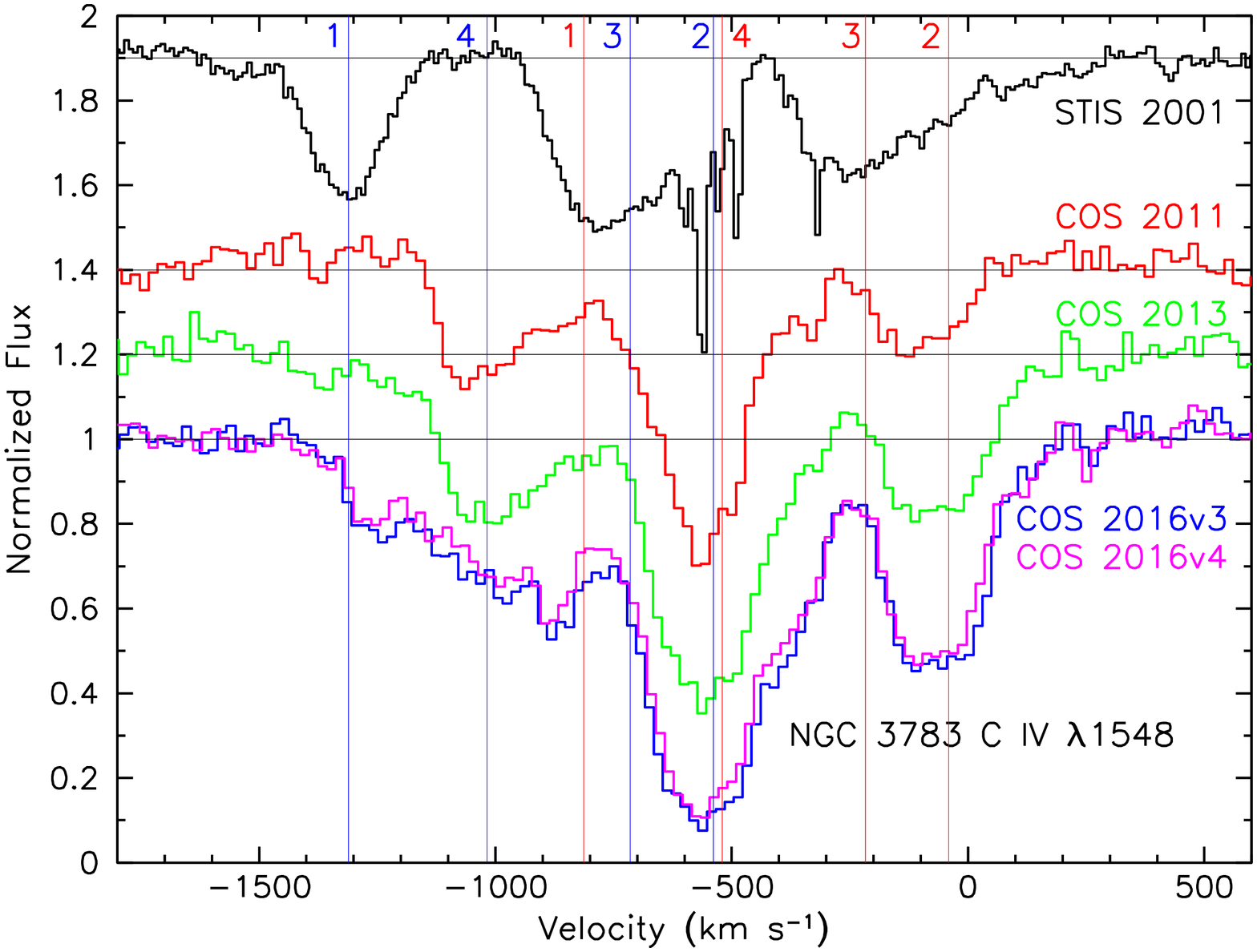}}
\caption{
Normalized HST spectra of the \ion{C}{iv} region from five different epochs
as labeled.
Velocity is for \ion{C}{iv} $\lambda$1548.195 relative to the host
galaxy systemic redshift of $z = 0.00973$ \citep{Theureau98}.
Fluxes are normalized to range from 0 to 1. Epochs prior to 2016 are offset
vertically by 0.2, 0.4, and 0.9, with the offset normalizations indicated
by thin horizontal black lines.
The velocities of components 1--4 as given by \cite{Gabel03b} are marked by
thin vertical blue lines and labeled; thin vertical red lines mark the expected
locations of the red component of the \ion{C}{iv} doublet.
}
\label{fig_c4_var}
\end{figure}

Our new observations in 2016 plus consideration of the intermediate epoch
spectrum in 2011 enables us to take a more comprehensive view.
In Figures \ref{fig_lya_var}, \ref{fig_n5_var}, and \ref{fig_c4_var}
we show normalized spectra at all four epochs (2001, 2011, 2013, and 2016)
for Ly$\alpha$, \ion{N}{v}, and \ion{C}{iv}.
The first thing to notice about the 2016 spectra is that all the absorption
features appear to be stronger and deeper.
At first this might seem puzzling since the UV continuum in 2016 is brighter
than during any of the other epochs. However, this is reminiscent of the
behavior of the narrow absorption lines in NGC 5548 during the obscuration
event starting in 2014 \citep{Kaastra14, Arav15, Goad16, Mathur17, Kriss18}.
Although the visible UV continuum in NGC 3783 is brighter, the soft X-ray
obscuration shows that over 70\% of the soft X-ray continuum is covered by
low-ionization, optically-thick gas. This would imply that only a small
percentage of the ionizing continuum is illuminating the narrow
absorption-line clouds. In fact, with the obscurer allowing only 26\%
transmission during the observation on 2016-12-12, the inferred ionizing
UV flux is comparable to the faint state during the 2013 observation.

\begin{table*}
  \caption[]{UV Continuum Fluxes in NGC~3783.
  }
  \label{tab:uvfluxes}
\begin{center}
\begin{tabular}{l c c c c c c c c}
\hline\hline
{Observation} & F(1470\AA)$^{\rm a}$ & F(912\AA)$^{\rm b}$ & $\rm T_f^c$ &  F(1470\AA)$^{\rm d}$ & F(912\AA)$^{\rm e}$ \\
              &            &           &           &  (inferred) & (inferred) \\
\hline
STIS 2001  &  3.76 & 16.4 & 1.00 & 8.47 & 16.4 \\
COS 2011   &  3.52 & 13.3 & 1.00 & 7.98 & 13.3 \\
COS 2013   &  1.53 & \phantom{0}5.9 & 1.00 & 3.51 & \phantom{0}5.9 \\
COS 2016-12-12 &  7.24 & 27.1 & 0.26 & 16.6 & \phantom{0}7.0 \\
COS 2016-12-21 &  7.97 & 30.2 & 0.32 & 18.6 & \phantom{0}9.7 \\
\hline
\end{tabular}
\end{center}
{\bf Notes.}\\
$^{\rm a}$ Observed continuum flux at 1470 \AA\ ($10^{-14} \rm~erg~cm^{-2}~s^{-1}~\AA$).\\
$^{\rm b}$ Fitted continuum flux extrapolated to 912 \AA\ and corrected for extinction assuming
$E(B-V)=0.107$ ($10^{-14} \rm~erg~cm^{-2}~s^{-1}~\AA$).\\
$^{\rm c}$ Fraction of the continuum transmitted by the obscurer \citep{Mehdipour17}.\\
$^{\rm d}$ Extinction-corrected continuum flux at 1470 \AA\ ($10^{-14} \rm~erg~cm^{-2}~s^{-1}~\AA$).\\
$^{\rm e}$ Inferred ionizing continuum flux at 912 \AA\ corrected for extinction and diminished by
transmission through the obscurer ($10^{-14} \rm~erg~cm^{-2}~s^{-1}~\AA$).\\
\end{table*}

To illuminate these differences in flux more quantitatively, we compare
the UV continuum fluxes at wavelengths both longward and shortward of
the Lyman limit at 912 \AA\ in Table \ref{tab:uvfluxes}.
We start with the observed UV continuum fluxes at 1470 \AA\ for
all observations, as measured directly from the spectra.
From the models we have fit to each spectrum, we extrapolate the
extinction-corrected power law down to the Lyman limit at 912 \AA.
The column labeled $\rm T_f$ then gives the fraction of light
transmitted by the obscurer for each observation, taken from the
partial covering models of the X-ray emission in \cite{Mehdipour17}.
In the next column we give the extinction-corrected flux at
1470 \AA.
In the last column, we assume that the fraction of light transmitted
by the obscurer is the same as that measured in the soft X-ray, and
then calculate the actual flux at 912 \AA\ after it is blocked by
the obscurer. Note that no obscurer was present in 2001, 2011, or 2013.

\begin{figure}[!tbp]
\centering
\resizebox{1.0\hsize}{!}{\includegraphics[angle=270]{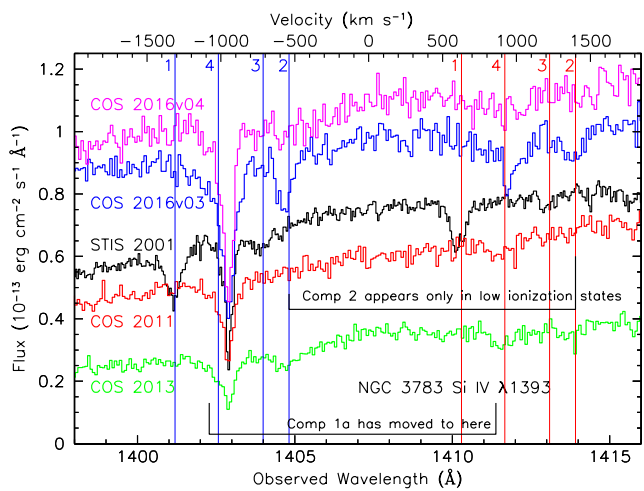}}
\caption{
HST spectra of the \ion{Si}{iv} region from five different epochs as labeled.
Fluxes and wavelengths are as observed.
The wavelengths of components 1--4 as given by \cite{Gabel03b} are marked by
thin vertical blue lines and labeled; thin vertical red lines mark the expected
locations of the red component of the \ion{Si}{iv} doublet.
Note that absorption in Component \#2, which is strong only in low-flux states
in all prior observations, appears strongest during the 2016-12-12 COS
observation, which was taken at the time that X-ray obscuration was strongest.
Also, the low-ionization Component \#1a, which is the ``decelerating"
absorption-line cloud \citep{Gabel03b, Scott14}, has decelerated to an observed
wavelength of 1403 \AA, where it is blended with
Galactic \ion{Si}{iv} $\lambda1403$.
}
\label{fig_si4_var}
\end{figure}

\begin{figure}[!tbp]
\centering
\resizebox{1.0\hsize}{!}{\includegraphics[angle=270]{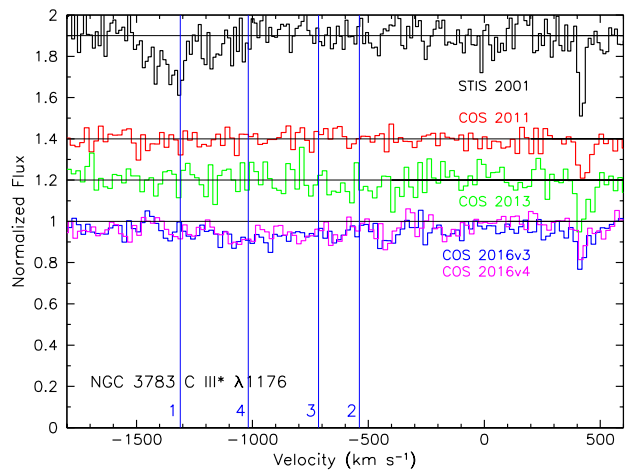}}
\caption{
Normalized HST spectra of the \ion{C}{iii}* $\lambda1176$
region from five different epochs as labeled.
Velocity is for the strongest \ion{C}{iii}* $J=2$ transition at 1175.71 \AA\ 
relative to the host galaxy
systemic redshift of $z = 0.00973$ \citep{Theureau98}.
Fluxes are normalized to range from 0 to 1. Epochs prior to 2016 are offset
vertically by 0.2, 0.4, and 0.9, with the offset normalizations indicated
by thin horizontal black lines.
The velocities of components 1--4 as given by \cite{Gabel03} are marked by
thin vertical blue lines and labeled.
}
\label{fig_c3_var}
\end{figure}

The consequences of the obscuration in NGC 3783
are most apparent in low-ionization lines such as
\ion{Si}{iv} (Fig. \ref{fig_si4_var}) and
\ion{C}{iii}* $\lambda 1176$  (Fig. \ref{fig_c3_var}).
Figure \ref{fig_si4_var} compares spectra of the \ion{Si}{iv} region
at all epochs in calibrated flux units.
During the 2001 campaign, \ion{Si}{iv} absorption appeared only in Component \#1,
most prominently during low-flux states \citep{Gabel05b}.
Note that there is no Component \#4 absorption in  \ion{Si}{iv} during 2001;
in fact, it is only noticeable as a strong feature in \ion{O}{vi}
in the {\it FUSE} spectrum \citep{Gabel03b}.
Component \#1 shifted in velocity gradually redward during the 2000--2001 campaign
\citep{Gabel03b}, and \cite{Scott14} suggest that the absorption appearing near
the velocity of Component \#4 in 2013 is actually the continued redward
evolution in velocity of Component \#1.
In the low-flux state of 2013, having ``moved" to the location of Component \#4,
it appears again.
Similarly, \ion{Si}{iv} absorption in Component \#2 only appears in low-flux states.
Likewise, \ion{C}{iii}* $\lambda 1176$ absorption was only associated with
Component \#1 in the 2000--2001 campaign, and was strongest in the low states
\citep{Gabel05b}.
As Fig. \ref{fig_c3_var} shows, \ion{C}{iii}* $\lambda 1176$ has reappeared
during the 2016 obscuration observations as a shallow depression near the
original velocity of Component \#4, but more likely representing the
evolution in velocity of gas associated with Component \#1.
As Table \ref{tab:uvfluxes} shows, the ionizing flux at 912 \AA\ in 2016 is
nearly as low as in 2013,
if most of the intrinsic continuum is hidden by the obscurer.
This shadowing of the narrow-absorption-line gas by the obscurer
then explains the appearance of absorption associated with
Components \#1 and \#2 during the obscuration event in 2016.

\subsubsection{Photoionization Response}

Several factors complicate the interpretation of the absorption-line profiles
in NGC 3783.
As \cite{Gabel03} showed in their analysis of the full Lyman series using
both the STIS and FUSE spectra, Ly$\alpha$ is heavily saturated. The absorption
lines are broad and blended.
In \ion{C}{iv}, the velocity spacings lead to overlap between the blue and red
lines of the \ion{C}{iv} doublet for Components \#1 and \#3, and for \#2 and \#4.
In \ion{N}{v} the individual troughs are only slightly blended, but
the red transition of Component \#4 is contaminated by foreground interstellar
absorption by \ion{S}{ii} $\lambda 1250$.
Since \ion{N}{v} has the cleanest profiles, we use it as a starting point for
all of our analysis.
In the three-year interval from 2013 to 2016, note that the figures show
little additional apparent motion of Component \#1.
In fact, its motion appears to have halted by the time of the 2011 spectrum,
where its location differs little from 2013 or 2016.
Another intriguing aspect of our 2016 spectra is that
absorption near the original position of Component \#1 has re-appeared.

Absorption at the original locations of Components \#2 and \#4 show a clear
ionization response, perhaps indicating that these are stable, persistent
features.
Absorption at the velocity of Component \#4 only appears in the low flux state
of the COS 2013 observation. Initially it may seem surprising that it is also
present in the first observation on 2016-12-12 (and not strongly in the
second), but this can be understood as another example of the line-of-sight
shadowing by the transient soft-X-ray obscuration, as
for the NGC 5548 obscurer \citep{Arav15}. If the UV flux is scaled down by the
transmission of the obscurer (only 26\% for the 2016-12-12 observation),
the inferred ionizing flux is actually {\it lower} than in 2013.
Similarly, in the 2016-12-21 observation, the obscuration is not as opaque,
the intrinsic continuum is slightly brighter, and the inferred ionizing
flux is {\it higher} than for the COS 2013 observation.

Similar arguments apply to the depth of the red side of the
absorption trough in Component \#2, although the interpretation is slightly
more complicated due to the different covering fractions of the line and
continuum for this feature, and the strongly varying intensity of the
emission lines relative to the continuum among the HST observations of
NGC 3783. Based on the analysis of the Lyman lines, \cite{Gabel03} show that 
the narrow absorption features have different covering factors for the
lines and the continuum. The depth of the troughs in Ly$\alpha$, which is
certainly saturated, for the 2016 observation show that over 92\% of
both line and continuum is covered.

To try to unravel the behavior of the narrow absorption components, we also
have examined their changing structure in the context of potential responses
to changes in the ionizing flux illuminating the absorbing clouds or
filaments. We start with the structures at the highest blue-shifted
velocities, Component \#1, then Component \#4, \#3, and \#2.

{\bf Component \#1.}
This component was strong and well defined in 2001 (and earlier). As described
by \cite{Gabel05b}, it appears to have both a
low-ionization component, \#1a, that appears in all ionic species,
including \ion{Si}{iv} and \ion{C}{iii}*.
It is blended with a high-ionization portion called
1b, which is needed to explain the \ion{N}{v} and \ion{O}{vi} strengths.
It is the low-ionization portion, \#1a, that appears to decelerate
\citep{Gabel03b}, although the shift in the line centroid is seen in all ions.
The most definitive detection is in our 2016-12-12 observation (v3),
where it appears in both \ion{Si}{iv} and \ion{C}{iii}*.
The \ion{C}{iii}* absorption associated with Component \#1 in 2016 appears as a
broad, shallow depression
at the velocity of Component \#4, visible in Fig. \ref{fig_c3_var}.

Similarly, the Ly$\alpha$ absorption formerly associated with Component \#1
in 2001 has also shifted to the position of Component \#4 in
the 2011, 2013, and 2016 spectra.
This new location for \#1a also shows an apparent response to ionizing
flux in that it is deepest for the 2013 spectrum, followed by
2016, and then 2011. This is consistent with higher neutral hydrogen
column densities during observations with lower {\it ionizing} flux
(i.e., observed UV flux corrected by transmission of the obscurer).

This apparent motion in Component \#1a seems to have separated it kinematically
from the high-ionization Component \#1b.
Just redward of the 2001 velocity of Component \#1, a new component
appears in 2011 in Ly$\alpha$ at $-1280~\rm km~s^{-1}$.
We suggest that this is the
counterpart to high-ionization Component \#1b. It also shows an
apparent response to changes in the ionizing flux, being deepest in
2013 and 2016, and shallower in 2011. Note that this velocity
is also more consistent with the blue side of the X-ray absorption
troughs in the high ionization ions, e.g., Figure 9 of \cite{Scott14},
comparing profiles for \ion{Mg}{xi} and \ion{Mg}{xii} to the UV ion \ion{N}{v}.

The behavior of this feature in \ion{N}{v} and \ion{C}{iv} corroborates
its identification as the high-ionization Component \#1b.
The feature stands out as well defined in \ion{N}{v} and \ion{C}{iv} in spectra
from 2016 when the ionizing continuum was obscured, but appears
only weakly in the prior spectra.
As shown by the photoionization models in Figure 7 of \cite{Gabel05b},
this is consistent with it having an ionization parameter log U $> -1.5$
(log $\xi > 0.0$), lying well beyond peak
ionization for \ion{N}{v} and \ion{C}{iv} during 2001 and 2011, but accumulating
higher column densities in those ions at the lower ionization
parameters more likely present during the obscured state.

{\bf Component \#4.}
As noted by \cite{Scott14}, the original Component \#1 appears to have moved
to the velocity of Component \#4.
In \ion{N}{v} this transition in velocity happened by the time of the
2011 observation. There is no discernible change in velocity
between 2011 and 2013, but the feature is much weaker in 2016.

In addition to this possible motion of Component \#1a, another trough
appeared redward of the Component \#4 location at $-880~\rm km~s^{-1}$ in 2011,
and this feature persists in 2013 and 2016. Its depth has no
relation to the strength of the ionizing flux--it is weakest in
2013 when the ionizing flux was weakest, stronger during the bright
2011 epoch, and at its strongest during the 2016 obscuration event.
However, in Ly$\alpha$ this feature shows a more consistent  response
to changes in the ionizing flux level. It is strongest in 2013
when the ionizing flux level was lowest, weakest in 2011 when it was
highest, and in between during the epoch of obscuration in 2016.

{\bf Component \#3.}
Component \#3 essentially disappears after 2001.
\cite{Scott14} note its absence in 2013, but it is also not present in the 2011
or the 2016 spectra.

{\bf Component \#2.}
Component \#2 was the most prominent absorber in Ly$\alpha$ in 2001, but it was
not exceptionally strong in any other ion.
Starting in 2011, it develops a profile in \ion{N}{v} and
\ion{C}{iv} more similar to its appearance in Ly$\alpha$.
Variations in Component \#2 are analyzed most cleanly in the \ion{N}{v}
transitions, where it is unblended.
It is saturated at all epochs in Ly$\alpha$, and it
is only visible in low-ionization epochs in \ion{Si}{iv} (2013 and 2016).
In \ion{C}{iv}, the blue component overlaps the red trough of Component \#4, or
the "decelerated" red trough of Component \#1a. The red trough of
Component \#2 is unblended. It therefore can be used to corroborate
inferences derived from analysis of the \ion{N}{v} doublet, but it cannot
provide an independent measure of covering fraction and optical depth.

The transition in the morphology of the Component \#2 absorption profile
from 2001 to 2011 is suggestive of (1) Component \#3 having decelerated from
its position in 2001 to form the blue side of the Component \#2 trough, and
(2) the red half of the trough corresponding to the original 2001 location of
Component \#2.
The evolution in strength of Component \#2 is consistent with a response to
changes in the ionizing flux. In 2001, when the ionizing UV flux was strongest,
Component \#2 was at its weakest.
During the obscured epoch of 2016, when the ionizing UV is weakest,
Component \#2 has its deepest troughs, with the \ion{N}{v} profile having an
appearance similar to the saturated Ly$\alpha$ profile.
The intermediate depths in 2011 and 2013 are in proportion to the relative
strengths of the UV flux at those epochs, both of which were fainter than 2001.
The red transition of Component \#2 in \ion{C}{iv} shows this same pattern
of changes in trough depth.
Since Ly$\alpha$ is strongly saturated, this suggests that this component is
highly ionized, as shown in Figure 7 of \cite{Gabel05b}, with the \ion{C}{iv}
and \ion{N}{v} ionization fractions being well past their peaks.

\section{Discussion}

Our simultaneous UV spectra of NGC 3783 quantify the broad, fast absorption
that appeared in the blue wings of the high-ionization UV resonance lines
at the same time as the appearance of strong soft X-ray obscuration
\citep{Mehdipour17}.
Our detailed analysis of the emission-line profiles show that broad absorption
on the blue wings of the permitted emission lines (and the absence of absorption in excited-state lines such as \ion{He}{ii}$\lambda1640$ and H$\beta$)
provides a more physically consistent description than an arbitrary set of
emission components.
Combining the spectral diagnostics of the obscurer
in our UV observations with the total
column density measured with the X-ray spectra enables us to determine
the ionization state (log $\xi = 1.84^{+0.4}_{-0.2}$ $\rm erg~cm~s^{-1}$)
and kinematics of the outflowing gas responsible for the obscuration.
Indeed, without the UV observations, it
would not even have been possible to assert that the soft X-ray obscuration
was due to an outflow since there are no prominent spectral features in the
heavily absorbed X-ray spectrum. Our UV spectra show that the absorption
extends from near zero velocity to a maximum of $\sim -6200~\rm km~s^{-1}$,
with a flux-weighted mean of $-2840~\rm km~s^{-1}$.

As in NGC 5548 \citep{Kaastra14}, the strength and depth of the UV
absorption varies in concert with variations in the X-ray obscuration.
When the obscuration is strongest, on 12-Dec-2016, the UV absorption
is strongest. As the obscuration lessened, as shown in the 21-Dec-2016
observation with {\it XMM-Newton}, the UV absorption also diminished.
This could be seen not only in the depths of the broad absorption features,
but also in its inferred influence on the ionizing UV continuum.
Low-ionization features in the intrinsic narrow absorption lines of
NGC 3783 such as \ion{C}{iii}* $\lambda$1176 and \ion{Si}{iv}
$\lambda\lambda$1393,1402 become deeper when the X-ray obscuration is strongest.
As in NGC 5548 \citep{Arav15}, this argues that the X-ray/UV obscurer lies
interior to the clouds producing the intrinsic narrow absorption lines.

Nearly two decades of high spectral resolution observations of
the narrow absorption-line features in NGC 3783
show that they are not as kinematically stable
as other AGN we have studied intensively, for example,
Mrk~509 \citep{Kriss11b, Arav12}, or NGC 5548 \citep{Arav15}.
The variations in the narrow absorption lines in NGC 3783 discussed in
\S\ref{narrow_line_var}
show that the behavior of these absorption features is not as simple as the
stable absorption troughs seen in NGC 5548.
In NGC 5548, these troughs maintain their velocities; they
show variations consistent with constant column density and covering factor
and simple changes in ionic column density determined purely by variations in
the ionizing flux. In NGC 3783, even for features that seem to have been stable
in velocity from 2011 to 2016, we see changes in both ionic column density
and in covering factor. Although some column-density changes are correlated with
changes in ionizing flux (e.g., Component \#1a and Component \#2), others are
not (Component \#1b, Component \#3, and Component \#4).
Thus, we must be viewing configurations of absorbing clouds that are clumpy,
and possibly
changing their arrangement along the line of sight. Given the relative
stability in velocity of some of the features, it seems that rather than having
clouds or filaments crossing our line of sight, or decelerating, that an
alternative possibility might be that the multiple velocities we see are
associated with dense clouds or filaments that are moving at fixed velocity.
These dense entities are a source of material, and, in and of themselves, they
have low enough covering fraction so as not to cause noticeable absorption.
The absorption we see is then caused by material ablated or evaporated from the
outer portions of these dense sources. This material is photoionized, heats up,
expands, and thus varies in covering fraction and ionization, giving rise to
the variable features we see.
Dense knots like this that form and evaporate can form naturally as thermal
instabilities in a radiatively accelerated flow \citep{Proga15}.

\begin{table}
  \caption[]{\ion{C}{iv} Emission-line Fluxes and Equivalent Widths in NGC~3783.
  }
  \label{tab:em_flux_history}
\begin{center}
\begin{tabular}{l c c c}
\hline\hline
{Observation} & F(1470\AA)$^{\rm a}$ & \ion{C}{iv} Flux$^{\rm b}$ &   \ion{C}{iv} EW$\rm ^c$ \\
\hline
STIS 2001 & 3.76 & 718    & 204  \\
COS 2011  & 3.52 & 676    & 200  \\
COS 2013  & 1.53 & 504    & 343  \\
COS 2016 (average) & 7.67 & 777    &  104 \\
\hline
\end{tabular}
\end{center}
{\bf Notes.}\\
$^{\rm a}$ Observed continuum flux at 1470 \AA\ ($10^{-14} \rm~erg~cm^{-2}~s^{-1}~\AA$).\\
$^{\rm b}$ Total observed \ion{C}{iv} emission-line flux (($10^{-14} \rm~erg~cm^{-2}~s^{-1}$).\\
$^{\rm c}$ \ion{C}{iv} emission-line equivalent width (EW) (\AA).\\
\end{table}

Rather than invoking deceleration to explain the changing kinematics of these
clumps, transverse motions of clouds along our line of sight
seems more plausible.
These transverse motions may be related to velocity shear in outflowing
streamlines that have both radial and toroidal motions defining their outflow
trajectories.
Based on the measured electron density and column density of Component \#1b,
\cite{Gabel05b} infer a transverse size of $10^{16}$ cm for a uniform cloud.
At the distance of 30 pc based on the photoionization solution,
Keplerian velocities at this location would be of order $200~\rm km~s^{-1}$,
which would include the gravitational potential of the host galaxy
as well as the black hole.
Thus it would take 2.1 years for the cloud to traverse our line of sight.
Since we see dramatic changes in 10 years or less, if these were caused by
transverse motion, the cloud would have to be elongated along our line of
sight by a factor of 10 relative to its width.
If ``clouds" are really filaments, perhaps organized along streamlines or
magnetic field lines (e.g., \cite{Fukumura10}), this is not implausible.

\subsection{Structure and Evolution of the Broad Line Region}

Based on the X-ray partial covering of the obscurer, which suggests its
transverse size is comparable to the size of the X-ray emitting continuum
region, and the variability timescale of $\sim 1$ day in the X-ray,
\cite{Mehdipour17} infer a density of $\sim2.6 \times 10^9~\rm cm^{-3}$ and
an approximate location of $\sim 10$ lt-days from the source.
The location, density, and the kinematics of the obscuring outflow all suggest
that it may be associated with the broad line region in NGC 3783.
As we showed in \S3, the broad emission line profiles have also changed
substantially between 2001 and 2016.

\begin{figure}[!tbp]
\centering
\resizebox{1.0\hsize}{!}{\includegraphics[angle=270]{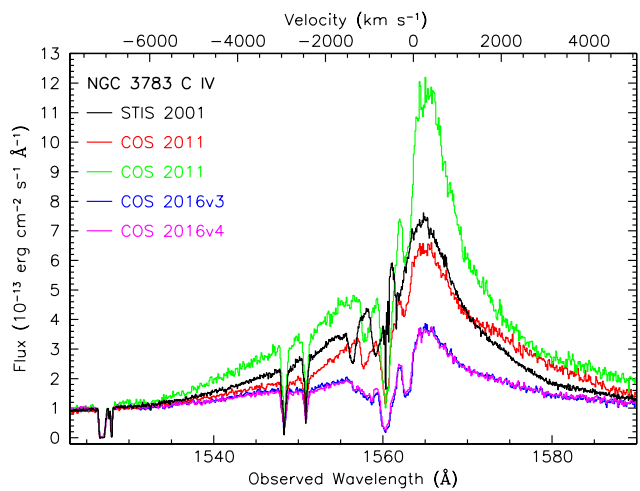}}
\caption{
Comparison of the \ion{C}{iv} emission line profiles for NGC 3783 in the
observations with
STIS 2001 (black), COS 2011 (red), COS 2013 (green), and COS 2016 Visit 3 (blue)
and COS 2016 Visit 4 (magenta).
The profiles have all been scaled to the flux level of COS 2016 Visit 4
at 1525 \AA.
}
\label{fig:c4_emission_profiles}
\end{figure}

\begin{figure}[!tbp]
\centering
\resizebox{1.0\hsize}{!}{\includegraphics[angle=270]{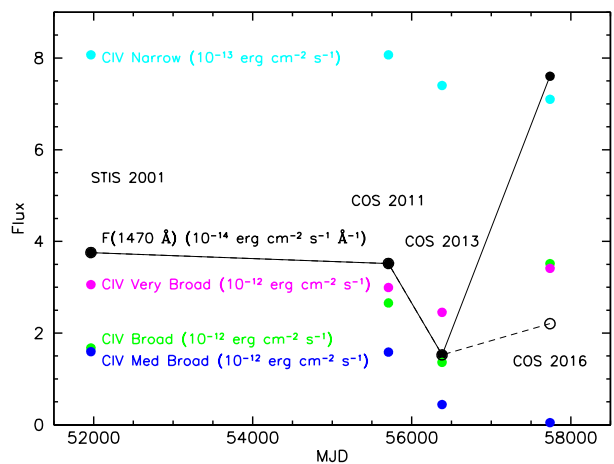}}
\caption{
Fluxes at the epochs of the STIS 2001, COS 2011, COS 2013, and COS 2016
observations of NGC 3783 in the continuum at 1470 \AA\ (black) and in the
several components of the \ion{C}{iv} emission line:
the narrow component (cyan),
the medium-broad component (green), the broad component (blue), and the
very broad component (magenta).
Fluxes are scaled as noted in the labels in the figure.
The continuum points are connected by a thin black line.
The open circle in 2016 shows the value of the continuum flux if we scale it
down by the transmission of the obscurer at the epoch of the COS observations.
}
\label{fig:emission_variations}
\end{figure}

The extensive reverberation mapping campaign conducted on NGC 3783 in 1992
by the International AGN Watch \citep{Reichert94, Stirpe94, Onken02}
showed that the UV emission lines respond to variations in
the continuum flux as expected if they are reprocessing the ionizing radiation
from the central source. Refined analysis of these data \citep{Onken02}
determined a mean lag for the  \ion{C}{iv} emission line of 3.8 days.
During the 1992 campaign, the line fluxes tracked the continuum closely, with
no dramatic changes in the line profile. However, as demonstrated in \S3,
the broad emission line profiles observed by COS in 2016 differ dramatically
from those seen with STIS in 2001.
The archival COS observations from 2011 and 2013 enables us to study the
evolution of these profiles. We have fit both of these spectra using the
same procedures as for the unobscured STIS spectrum.
The best-fit parameters for each are tabulated in Table \ref{tab:cos11cos13_em}.
To show a more comprehensive history of
these variations, we compare the  \ion{C}{iv} emission line profiles for
all four epochs from 2001 to 2016 in Figure \ref{fig:c4_emission_profiles}.
The most striking aspect of this comparison
is the large variation in equivalent width among these observations.
Table \ref{tab:em_flux_history} summarizes the total fluxes and equivalent
widths (EW) of the \ion{C}{iv} emission lines at all four epochs in our study.
Note that although both the continuum and the \ion{C}{iv} flux were at a maximum
in 2016, the equivalent width was at a minimum.
Conversely, when flux was at a minimum in 2013, the equivalent width was at a
maximum. The narrow core of the line dominates the emission profile (at
velocities $< 500~\rm km~s^{-1}$) in 2013, but it makes only a minor
contribution in 2016.
In fact, the narrow component stays relatively constant in flux,
similar to the narrow core in NGC 5548 \citep{Crenshaw09}.

Figure \ref{fig:emission_variations} shows these variations quantitatively.
The very broad component varies roughly in proportion to the variations
in the continuum flux.
The medium-broad (FWHM$\sim$2838 $\rm km~s^{-1}$) component shows the most
variation, virtually disappearing during the COS observations of 2016.
This same kinematic component is also much weaker in the \ion{N}{v},
\ion{Si}{iv}, \ion{He}{ii}, and H$\beta$ line profiles.\footnote{
The medium-broad component of Ly$\alpha$ does not share this behavior.
Strong absorption due to damped Milky Way Ly$\alpha$ absorption blends with
all components on the blue wing of Ly$\alpha$ in NGC 3783 and adds
considerable degeneracy to our decomposition of the line profile.
}
Although our decomposition of the line profile is not unique, given the
kinematic relationships between line width and reverberation lags
established in many reverberation campaigns, we can roughly decompose the
\ion{C}{iv} emission-line profile into a correspondence between width and
distance from the central source using the corresponding Keplerian orbital
velocities in the potential of the
$2.35 \times 10^7~\Msun$ black hole \citep{Bentz15} in NGC 3783.
The Keplerian velocity corresponding to the mean lag of 3.8 lt-days
is $5630~\rm km~s^{-1}$.
For our model of the  \ion{C}{iv} profile,
the narrow core ($937~\rm km~s^{-1}$) lies at a distance of 137 lt-days,
the medium-broad component ($2838~\rm km~s^{-1}$) at 15 lt-days,
the broad component ($4576~\rm km~s^{-1}$) at 5.8 lt-days,
and the very broad component ($10030 ~\rm km~s^{-1}$) at 1.2 lt-days.

\begin{figure}[!tbp]
\centering
\resizebox{1.0\hsize}{!}{\includegraphics[angle=0]{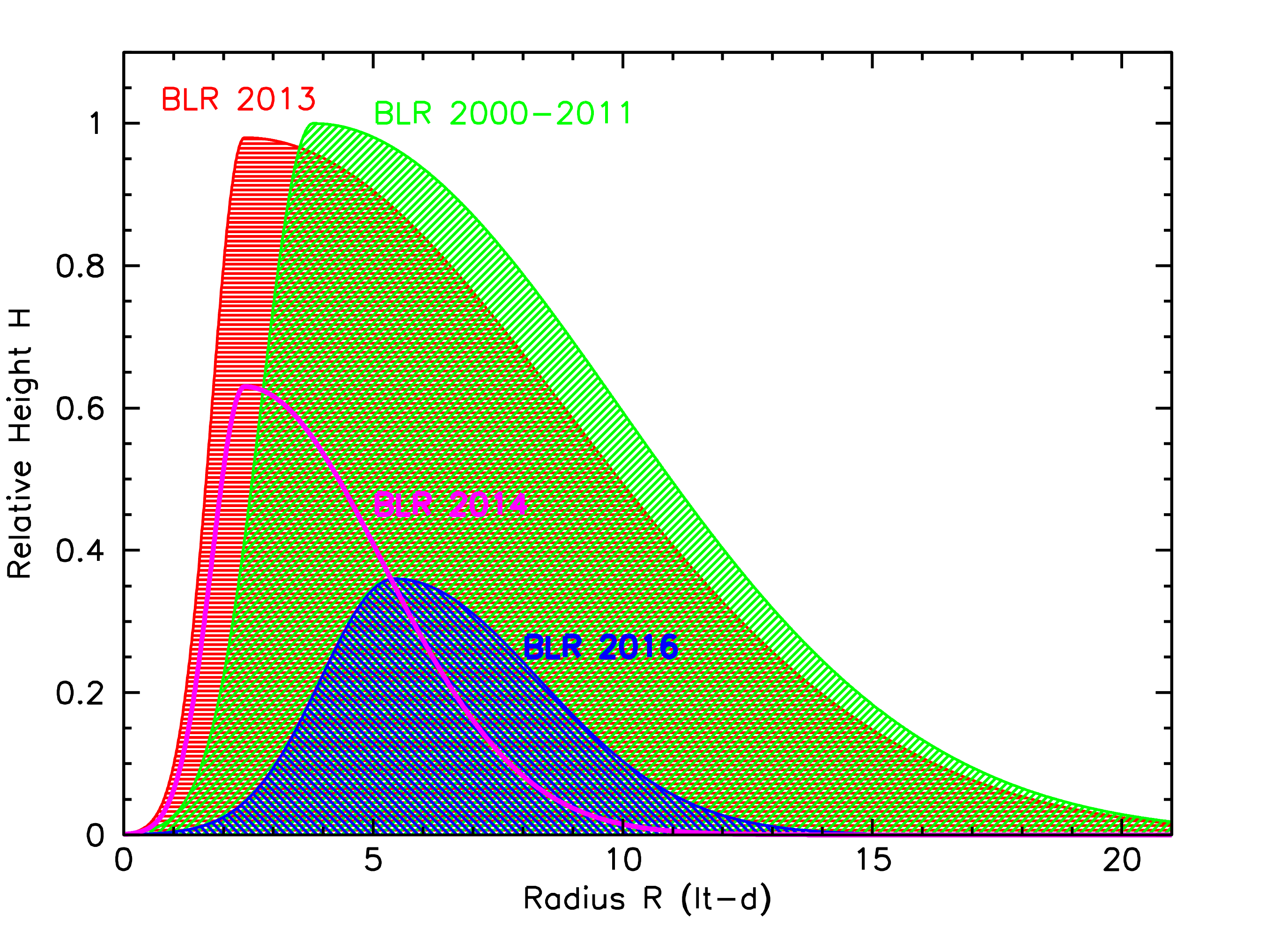}}
\caption{
Schematic representation of the evolution of the BLR in NGC 3783
The green shaded area represents the vertical structure and radial extent of
the BLR during the period from 2000 to 2011.
The red shaded area shows the inferred structure in 2013, when the continuum
flux was 2.5x less, but the EW of \ion{C}{iv} was 70\% higher.
The magenta curve is the hypothetical structure in 2014, when the continuum
flux was still low after the BLR had dynamically adjusted to the lower flux
level.
The blue shaded area is the inferred structure at the time of the
obscuration event in 2016.
}
\label{fig:blr_evolution}
\end{figure}

In the context of our model of the line profile, both 2013 and 2016 represent
unusual states for the broad line region of NGC 3783.
Overall, the continuum in 2013 is 2.5--5x fainter than in
2001 or 2016, yet the very broad component of the \ion{C}{iv} emission-line flux
is dimmer by only 20--30\%.
If the BLR is reprocessing continuum radiation, to compensate for the reduced
continuum flux and still radiate nearly as profusely, the covering factor
of the very broad portion of the \ion{C}{iv} profile must be higher.
This implies that in the faint state of 2013, the inner part of the BLR must
have puffed up vertically.
At the same time, we note that the medium-broad component at
$2838~\rm km~s^{-1}$ virtually disappeared in 2016.

We can interpret these changes in the context of recent models in
which a dust- and line-driven wind from the accretion disk create the BLR
\citep{Czerny11, Czerny17, Baskin18}.
Although these models do not yet quantitatively predict sizes and velocities
for the BLR that match the emission-line profiles and velocities observed
in typical AGN \citep{Czerny17},
they offer a physical framework that may explain some of the qualitative
changes we observed in the spectra of NGC 3783 over the years.
We hypothesize that the dramatic drop in continuum flux in 2013 (or shortly
before) initiated a reconfiguration of the BLR in NGC 3783.
Figure \ref{fig:blr_evolution} illustrates this evolution schematically.

From the 1992 reverberation-mapping campaign through the STIS observations in
2000--2001 and the COS observation in 2011, the broad-line profiles in
NGC 3783 remained similar in shape and equivalent width.
This apparently stable configuration is represented in
Figure \ref{fig:blr_evolution} by the shaded green area.
The shape is chosen to resemble line emissivity distributions typical of
reconstructions from reverberation-mapping results \citep{Krolik91, Grier13}.
The peak is set at the \ion{C}{iv} mean lag of 3.8 lt-days.
To scale this configuration to other epochs in the history of NGC 3783
observations, we use the scaling relations of \cite{Baskin18} for a dusty
wind-driven BLR.
The peak radius of the broad line region, $\rm R_{BLR}$,
scales with luminosity, L, as $\rm R_{BLR} \sim L^{1/2}$.
The height of the broad line region, H,
also scales with luminosity as $\rm H \sim L^{1/2}$.
Thus, when the continuum brightness drops in 2013, the peak of the BLR
emissivity shifts to a lower radius, R=2.4 lt-days.
Of course, the BLR can not change its shape instantaneously. For the
radiation-pressure driven winds in the \cite{Czerny17} and \cite{Baskin18}
models, the relevant timescale is the local dynamical timescale for material
moving up or down from the accretion disk, set by the vertical component of
the gravitational force exerted by the central black hole.
The free-fall time for material falling back to the disk after the radiation
pressure disappears is
$$\rm t_{ff} = \left(\frac{2 R^3}{GM_{BH}}\right)^{1/2} .$$

So, in 2013, the location of peak emissivity is expected to move inward 
to 2.4 lt-days in response to the reduced brightness of the central source and
the accretion disk.
The expected BLR height at that location would also be expected to be lower,
at $\sim64$\% of the peak height in 2000-2011.
However, it is likely that highly ionized gas above the disk would drop in
ionization very quickly, producing BLR-like emissivity at large heights.
The red shaded area in Figure \ref{fig:blr_evolution} reflects the smaller
$\rm R_{BLR}$, but a height consistent with the higher observed equivalent
width of the \ion{C}{iv} emission line.
Gas at these heights either at small radii in the disk or at larger radii
can not be supported via radiation pressure, but the dynamical adjustments in
the vertical extent of the BLR would take months
to years to adjust to this lower luminosity.
At 2.4 lt-days, the timescale is $\rm t_{ff} = 0.3$ years;
at 10 lt-days, the peak radius for H$\beta$ in the 1992 reverberation campaign,
it is 3.4 years.
The red shaded area in Figure \ref{fig:blr_evolution} reflects these
delays. We do not know when the drop in continuum flux actually occurred
prior to the 2013 COS observation, but the large equivalent width
observed in the emission lines
implies that gas at those innermost radii has not yet had time to fall back
toward the disk. Therefore we infer that the drop in flux happened very soon
before the 2013 observation. Similarly, since the gas at larger radii would take
years to fall back, the red shaded area shows a similar height to the prior
shape represented by the green shaded area.

We have no observations of NGC 3783 in 2014, so the magenta curve in
Figure \ref{fig:blr_evolution} is a hypothetical view of how the BLR would
have responded dynamically to the lower flux observed in 2013, had it remained
at that low flux level. By then, gas at heights supported by the normal flux
state in 2000--2011 would have had time to fall back toward the accretion disk.
The peak height of the BLR is adjusted by the square root of the flux ratios
between 2001 and 2013, and the vertical extent of the BLR at all radii
drops significantly due to the loss of radiation-pressure support.

{\it Swift} observations of NGC 3783 \citep{Kaastra18} show that NGC 3783
was still in a low flux state in mid-to-late 2016, several months before
the obscuration event in late 2016 \citep{Mehdipour17}.
Therefore, even though NGC 3783 had brightened dramatically, gas in the BLR
would not have had time to move in response to these changes.
The blue shaded area in Figure \ref{fig:blr_evolution} shows the peak
of the BLR emissivity moving to a larger radius, 5.4 lt-days, but not changing
much in vertical height since it has not yet had time to
respond dynamically to the increased radiation pressure.
The height shown in Figure \ref{fig:blr_evolution} is scaled to give the
observed ratio of \ion{C}{iv} equivalent width in 2016 to that in 2000--2001.
This height, however, is about a factor of 4x lower than the increased flux
in 2016 could potentially support via radiation pressure.
Thus, gas at all radii is likely flowing upward and outward from the plane of
the accretion disk as it is accelerated by the increased radiative flux.
In addition, there is a substantial amount of gas at radii interior to 5 lt-days
at high elevations that is now exposed to a much stronger ionizing flux.
We suggest that this gas and the outflow induced by the increased ionizing
flux is the source of the obscuring outflow observed by \cite{Mehdipour17}.

The changes we have observed in the structure of the BLR in NGC 3783,
and perhaps in the mechanisms driving these changes, are similar to the more
extreme variability seen in ``changing-look" AGN.
These AGN at their most extreme appear to change from Type 1, with bright,
typical BLRs, to Type 2, with almost no remaining observational trace of the
BLR.
Examples date back to the earliest studies of AGN
\citep[e.g., Mrk 6,][]{Khachikian71} and NGC 4151 \citep{Penston84}.
More recent, intensively studied examples include Mrk 590
\citep{Denney14, Mathur18} and HE1136-2304 \citep{Zetzl18}.
In such objects, the continuum luminosity varies by one to two orders of
magnitude; in the low-luminosity states, the BLR can nearly completely
disappear, the extreme version of the dimunition we observed in NGC 3783.
In models of the BLR produced by disk-driven winds, the BLR is expected to
 completely disappear below a critical luminosity required to sustain the
outflow \citep{Elitzur09, Elitzur14}, producing an AGN that is a ``true Type 2",
where the BLR is not merely obscured, but completely absent.
\cite{Elitzur09} cite a threshold for such a transition at a bolometric
luminosity of
$L_{bol} = 1.5 \times 10^{38} (M_{BH}/10^7 \Msun)^{2/3}~\rm erg~s^{-1}$.
For the black hole mass of $2.35 \times 10^7~\Msun$ in NGC 3783, this
threshold is far below the bolometric luminosity of
$2.8 \times 10^{44}~\rm erg~s^{-1}$
\citep[based on the spectral energy distributions of][]{Mehdipour17},
so we would not expect such a drastic reconfiguration of the BLR.

Comprehensive surveys of candidates in the Sloan Digital Sky Survey
(SDSS) reveal that 30--50\% of quasars exhibit extreme variability
characteristic of changing-look AGN that lead to dramatic changes in the BLR
\citep{MacLeod16, Rumbaugh17, MacLeod18}.
These objects have systematically lower Eddington ratios ($<$10\%), and
are similar in luminosity and Eddington ratio to Seyfert 1s like NGC 3783
$(L_{bol} / L_{Edd} \sim 0.02)$.
Thus, detailed study of a less extreme event such as we have observed may
offer some insights into the physics governing AGN classification in general.
A crucial difference between the events in NGC 3783 and in the changing-look
quasars is the appearance of heavy obscuration accompanying the
return to normal luminosities in NGC 3783.
This may imply that in changing-look objects that much of the material in
the accretion disk and the accompanying gas that supplies the wind for the
BLR may have been emptied from the system during the low-luminosity state.
This would leave no excess material
to be ejected as the obscuring outflow during the return to normal brightness
in the changing-look AGN.

\subsection{Impact on the Host Galaxy}

A high velocity wind originating in or near the BLR in an AGN can potentially
have a significant impact on the host galaxy since the kinetic luminosity of an
outflow varies with velocity, v, as $\rm v^3$.
For a spherical shell moving at this velocity with a radius R, a total
hydrogen column density $\rm N_H$ and covering fraction $\Delta\Omega$,
the mass flux and kinetic luminosity of the outflow are
{\center
\vskip -20pt
$$\rm \dot M = 4\pi \Delta\Omega R N_H \mu m_p {\rm v} $$
$$ \rm \dot E_k = \frac{1}{2} \dot M {\rm v}^2 $$
}
\noindent
where $\rm m_p$ is the proton mass and
$\mu = 1.4$ is the mean molecular weight.
The properties of the obscurer in NGC 3783 suggest it might have sufficient
kinetic luminosity to have an evolutionary impact on the host galaxy.

\cite{Mehdipour17} derive a radial distance for the obscurer of 10 lt-days;
the \ion{C}{iv} emission line region has a reverberation radius of 3.8 lt-days
\citep{Reichert94}.
For a central black hole mass of $\rm M_{BH} = 2.35 \times 10^7~\Msun$
\citep{Bentz15}, Keplerian velocities at 10 lt-days and 3.8 lt-days are
3500 $\rm km~s^{-1}$ and 5600 $\rm km~s^{-1}$, respectively.
These are typical of the velocities of the deepest point in the UV absorption
troughs of the obscurer ($-3000~\rm km~s^{-1}$), and their highest velocity
extent from line center at $-6200~\rm km~s^{-1}$, which suggest that the
kinematics of the obscurer are similar to those of the BLR.
Assuming a covering fraction of $\Delta\Omega=0.25$,
these velocities yield kinetic luminosities ranging from
$9.4 \times 10^{41}~\rm erg~s^{-1}$ to
$3.8 \times 10^{42}~\rm erg~s^{-1}$.
For the bolometric luminosity during our December 2016 observations
of $2.8 \times 10^{44}~\rm erg~s^{-1}$ \citep{Mehdipour17},
and an Eddington luminosity of $2.8 \times 10^{45}~\rm erg~s^{-1}$
(for $\rm M_{BH} = 2.35 \times 10^7~\Msun$),
the kinetic luminosity corresponds to a range of 0.34\% to 1.4\% of
bolometric, but only 0.03--0.14\% of the Eddington luminosity.
Even if this high-velocity outflow is persistent, these kinetic luminosities
are not quite sufficient to have an evolutionary impact on the host galaxy,
for which they generally must lie in the range of 0.5--5\%
\citep{DiMatteo05,Hopkins10} of the AGN Eddington luminosity.

\section{Conclusions}

Our UV spectra of NGC 3783 obtained with {\it HST}/COS in simultaneity with
{\it XMM-Newton} X-ray observations triggered with {\it Swift} by a soft X-ray
obscuration event reveal that the obscurer is a fast, broad outflow similar
in character to those recently discovered in
NGC 5548 \citep{Kaastra14}, Mrk 335 \citep{Longinotti13}, and
NGC 985 \citep{Ebrero16}.
The outflow extends from zero velocity to a maximum velocity
of $-6200~\rm km~s^{-1}$, with a maximum depth at $-3000~\rm km~s^{-1}$,
and a mean depth at $-2840~\rm km~s^{-1}$.
The UV absorption is visible in Ly$\alpha$, \ion{N}{v}, \ion{Si}{iv}, and
\ion{C}{iv}, indicating a moderate level of ionization.
The X-ray determined column density of $N_H=2.3 \times 10^{23}~\rm cm^{-2}$
together with the UV ion column densities yield an ionization parameter of
log $\xi =1.84^{+0.4}_{-0.2}$ $\rm erg~cm~s^{-1}$,
similar to highly ionized portions of the broad-line region.

Even though NGC 3783 was in an historically high UV flux state, the intrinsic
absorption lines exhibit depths and ionization states similar to the lowest
flux states seen in the past. The appearance of deep troughs and low-ionization
states like \ion{C}{iii} demonstrate that despite the high UV continuum flux,
the ionizing UV must be shadowed by the soft X-ray obscurer, similar to the
behavior seen in NGC 5548 \citep{Arav15}.

The broad emission line profiles in NGC 3783 changed dramatically as well.
Despite the high UV continuum flux, the emission line fluxes are comparable
to those seen with STIS in 2001, and moderate-width ($\sim 2800~\rm km~s^{-1}$)
portions of the broad-line profile have disappeared.
We suggest that the central portions of the broad line region have collapsed in
an infall, triggering the brightening observed in the UV and X-ray, and also
triggering the obscuring outflow. Our observations illustrate the value of
simultaneous UV and X-ray spectral observations to understand the kinematics
and physical properties of obscuring events.

\begin{acknowledgements}
This work is based on observations obtained with 
the NASA/ESA HST, and obtained from the Hubble
Legacy Archive. This work was supported by NASA
through a grant for HST program number 14481 from the Space Telescope Science
Institute, which is operated by the Association of Universities for Research in
Astronomy, Incorporated, under NASA contract NAS5-26555.
We also used XMM-Newton, an ESA science mission with instruments and
contributions directly funded by ESA Member States and 
the USA (NASA), the NuSTAR mission,
a project led by the California Institute of Technology (Caltech), managed 
by the Jet Propulsion Laboratory (JPL) and funded by NASA, and data supplied by
the UK Swift Science Data Centre at the University of Leicester.
We thank the {\it XMM-Newton}, {\it NuSTAR}, and HST teams for scheduling our
Target of Opportunity triggered observations.
SRON is supported financially by NWO, the Netherlands Organization for
Scientific Research.
SB acknowledges financial support from the Italian Space Agency under grant
ASI-INAF I/037/12/0, and n. 2017-14-H.O.
EC is partially supported by the NWO-Vidi grant number 633.042.525.
BDM acknowledges support from the European Union's Horizon 2020 research and
innovation programme under the Marie Sk{\l}odowska-Curie grant agreement
No. 665778 via the Polish National Science Center grant
Polonez UMO-2016/21/P/ST9/04025.
POP acknowledges support from the CNES and CNRS/PNHE.
GP acknowledges support from the Bundesministerium f\"ur Wirtschaft und
Technologie/Deutsches Zentrum f\"ur Luft- und Raumfahrt
(BMWI/DLR, FKZ 50 OR 1604) and the Max Planck Society.
\end{acknowledgements}

\vspace{-0.8cm}
\bibliographystyle{aa}
\bibliography{ngc3783}

\begin{sidewaystable*}
\begin{center}
        \caption[]{Emission-Line Parameters for the COS 2011 and COS 2013 Spectra of NGC~3783}
        \label{tab:cos11cos13_em}
\begin{tabular}{l c c c c c c c c}
\hline\hline
        &      & \multicolumn{3}{c}{COS 2011} &   &  \multicolumn{3}{c}{COS 2013}   \\
Feature & $\rm \lambda_0^{\mathrm{a}}$ &  $\rm Flux^{\mathrm{b}}$ & $\rm v_{sys}^{\mathrm{c}}$ & $\rm FWHM^{\mathrm{d}}$ & \phantom{000} & $\rm Flux^{\mathrm{b}}$ & $\rm v_{sys}^{\mathrm{c}}$ & $\rm FWHM^{\mathrm{d}}$ \\
\hline
\ion{C}{iii}\*         &   1176.01 & $   1.1\pm  0.2$ & $  -100 \pm  80$ & $ 1200 \pm  180$ & \phantom{000} & $   1.5\pm  0.3$ & $     0 \pm  50$ & $ 1200 \pm  260$\\
Ly$\alpha$             &   1215.67 & $ 110.0\pm  3.7$ & $  -110 \pm  20$ & $  800 \pm   20$ & \phantom{000} & $ 100.0\pm  3.1$ & $   -40 \pm  10$ & $  800 \pm   20$\\
Ly$\alpha$             &   1215.67 & $ 150.0\pm  5.9$ & $   180 \pm  20$ & $ 2430 \pm   30$ & \phantom{000} & $ 170.0\pm  5.8$ & $    90 \pm  10$ & $ 2310 \pm   10$\\
Ly$\alpha$             &   1215.67 & $ 260.0\pm  9.3$ & $   890 \pm  60$ & $ 5190 \pm   60$ & \phantom{000} & $  77.0\pm  3.9$ & $  1760 \pm  40$ & $ 6070 \pm   50$\\
Ly$\alpha$             &   1215.67 & $ 230.0\pm  7.6$ & $ -1160 \pm  60$ & $14870 \pm   50$ & \phantom{000} & $ 160.0\pm  5.3$ & $   200 \pm  20$ & $10590 \pm   30$\\
\ion{N}{v} blue        &   1238.82 & $   5.1\pm  0.3$ & $   -30 \pm  30$ & $  770 \pm   80$ & \phantom{000} & $   7.4\pm  0.3$ & $   -30 \pm  10$ & $  770 \pm   20$\\
\ion{N}{v} red         &   1242.80 & $   5.1\pm  0.3$ & $   -30 \pm  30$ & $  770 \pm   80$ & \phantom{000} & $   7.4\pm  0.3$ & $   -30 \pm  10$ & $  770 \pm   20$\\
\ion{N}{v} blue        &   1238.82 & $   6.0\pm  0.4$ & $  1080 \pm  20$ & $ 4240 \pm   60$ & \phantom{000} & $  11.0\pm  0.5$ & $   580 \pm  20$ & $ 3690 \pm   20$\\
\ion{N}{v} red         &   1242.80 & $   6.0\pm  0.4$ & $  1080 \pm  20$ & $ 4240 \pm   60$ & \phantom{000} & $  11.0\pm  0.5$ & $   580 \pm  20$ & $ 3690 \pm   20$\\
\ion{N}{v}             &   1240.89 & $  89.0\pm  3.0$ & $   480 \pm  20$ & $ 7270 \pm   60$ & \phantom{000} & $   2.8\pm  0.4$ & $   450 \pm 930$ & $13100 \pm  720$\\
\ion{S}{ii}            &   1260.42 & $   1.1\pm  0.5$ & $  -150 \pm  20$ & $ 1600 \pm   40$ & \phantom{000} & $   6.4\pm  0.4$ & $  -150 \pm  30$ & $ 2460 \pm   30$\\
\ion{O}{i}+\ion{S}{ii} &   1304.46 & $  11.0\pm  0.8$ & $   220 \pm  20$ & $ 2880 \pm  100$ & \phantom{000} & $  11.0\pm  0.4$ & $   220 \pm  20$ & $ 2920 \pm   30$\\
\ion{C}{ii}            &   1334.53 & $   4.3\pm  0.4$ & $   960 \pm  50$ & $ 2880 \pm  100$ & \phantom{000} & $   4.0\pm  0.2$ & $   110 \pm  30$ & $ 2920 \pm   30$\\
\ion{S}{iv} blue       &   1393.76 & $   4.3\pm  0.4$ & $  -280 \pm  30$ & $ 1310 \pm  100$ & \phantom{000} & $   3.6\pm  0.2$ & $   -70 \pm  50$ & $ 1260 \pm   30$\\
\ion{S}{iv} red        &   1402.77 & $   4.3\pm  0.4$ & $  -510 \pm  90$ & $ 1310 \pm   20$ & \phantom{000} & $   3.6\pm  0.2$ & $   -70 \pm  50$ & $ 1260 \pm   30$\\
\ion{S}{iv} blue       &   1393.76 & $  19.0\pm  0.6$ & $  1060 \pm  20$ & $ 4740 \pm   50$ & \phantom{000} & $  23.0\pm  1.3$ & $  -120 \pm  10$ & $ 5920 \pm  100$\\
\ion{S}{iv} red        &   1402.77 & $  19.0\pm  0.6$ & $  1060 \pm  20$ & $ 4740 \pm   50$ & \phantom{000} & $  23.0\pm  1.3$ & $  -120 \pm  10$ & $ 5920 \pm  100$\\
\ion{S}{iv}            &   1398.19 & $  42.0\pm  1.4$ & $  -360 \pm  50$ & $10850 \pm   50$ & \phantom{000} & $  15.0\pm  1.2$ & $ -3060 \pm  70$ & $17800 \pm   50$\\
\ion{O}{iv}]           &   1401.16 & $   1.8\pm  0.1$ & $   700 \pm  40$ & $ 1400 \pm  130$ & \phantom{000} & $   2.1\pm  0.3$ & $   900 \pm  40$ & $ 1260 \pm   10$\\
\ion{O}{iv}]           &   1401.16 & $  11.0\pm  0.6$ & $   700 \pm  40$ & $ 4740 \pm   20$ & \phantom{000} & $   1.9\pm  0.4$ & $   900 \pm  40$ & $ 5920 \pm   10$\\
\ion{N}{iv}]           &   1486.50 & $   0.0\pm  0.5$ & $    10 \pm  30$ & $ 5000 \pm  190$ & \phantom{000} & $   0.7\pm  0.4$ & $    10 \pm  20$ & $  500 \pm   20$\\
\ion{N}{iv}]           &   1486.50 & $  12.0\pm  0.5$ & $  -260 \pm  20$ & $ 2590 \pm   60$ & \phantom{000} & $   9.5\pm  0.4$ & $  -260 \pm  10$ & $ 2590 \pm   70$\\
\ion{C}{iv} blue       &   1548.19 & $  40.0\pm  1.5$ & $   -60 \pm  20$ & $  940 \pm   20$ & \phantom{000} & $  37.0\pm  1.6$ & $   -60 \pm  20$ & $  940 \pm   10$\\
\ion{C}{iv} red        &   1550.77 & $  40.0\pm  1.5$ & $   -60 \pm  20$ & $  940 \pm   20$ & \phantom{000} & $  37.0\pm  1.6$ & $   -60 \pm  20$ & $  940 \pm   10$\\
\ion{C}{iv} blue       &   1548.19 & $   7.9\pm  0.8$ & $   300 \pm  80$ & $ 2840 \pm   50$ & \phantom{000} & $  22.0\pm  1.6$ & $   500 \pm  30$ & $ 1390 \pm   70$\\
\ion{C}{iv} red        &   1550.77 & $   7.9\pm  0.8$ & $   300 \pm  80$ & $ 2840 \pm   50$ & \phantom{000} & $  22.0\pm  1.6$ & $   500 \pm  30$ & $ 1390 \pm   70$\\
\ion{C}{iv} blue       &   1548.19 & $ 140.0\pm  4.7$ & $   500 \pm  20$ & $ 4580 \pm   20$ & \phantom{000} & $  68.0\pm  2.3$ & $  -620 \pm  30$ & $ 5030 \pm   40$\\
\ion{C}{iv} red        &   1550.77 & $ 140.0\pm  4.7$ & $   500 \pm  20$ & $ 4580 \pm   20$ & \phantom{000} & $  68.0\pm  2.3$ & $  -620 \pm  30$ & $ 5030 \pm   40$\\
\ion{C}{iv}            &   1549.48 & $ 300.0\pm 10.0$ & $  -430 \pm  20$ & $ 9840 \pm   30$ & \phantom{000} & $ 250.0\pm  7.9$ & $  -140 \pm  10$ & $ 9220 \pm   20$\\
\ion{He}{ii}           &   1640.45 & $   9.1\pm  0.6$ & $   -80 \pm  40$ & $  790 \pm   50$ & \phantom{000} & $   6.9\pm  0.3$ & $    90 \pm  20$ & $  790 \pm   40$\\
\ion{He}{ii}           &   1640.45 & $  15.0\pm  0.6$ & $   840 \pm  90$ & $ 3250 \pm   30$ & \phantom{000} & $  19.0\pm  1.1$ & $  -110 \pm  30$ & $ 3450 \pm  120$\\
\ion{He}{ii}           &   1640.45 & $  98.0\pm  3.0$ & $  -120 \pm  20$ & $11080 \pm   30$ & \phantom{000} & $  39.0\pm  1.9$ & $ -5330 \pm  40$ & $23200 \pm  660$\\
\ion{O}{iii}]          &   1660.81 & $   5.5\pm  0.3$ & $  -160 \pm  20$ & $ 1200 \pm   20$ & \phantom{000} & $   3.5\pm  0.4$ & $  -160 \pm  20$ & $ 2790 \pm   60$\\
\ion{O}{iii}]          &   1666.15 & $   6.5\pm  0.3$ & $  -160 \pm  20$ & $ 1200 \pm   20$ & \phantom{000} & $  15.0\pm  0.7$ & $  -160 \pm  10$ & $ 2790 \pm   60$\\
\ion{N}{iii}]          &   1750.00 & $  13.0\pm  0.4$ & $     0 \pm  30$ & $ 3270 \pm   20$ & \phantom{000} & $  15.0\pm  0.5$ & $     0 \pm  40$ & $ 3270 \pm   30$\\
\hline
\end{tabular}
\end{center}
{\bf Notes.}\\
$^{\mathrm{a}}$ Vacuum rest wavelength of the spectral feature (\AA).\\
$^{\mathrm{b}}$ Integrated flux in units of $\rm 10^{-14}~erg~cm^{-2}~s^{-1}$.\\
$^{\mathrm{c}}$ Velocity (in $\rm km~s^{-1}$) relative to a systemic redshift of z = 0.00973 \citep{Theureau98}.\\
$^{\mathrm{d}}$ Full-width at half-maximum ($\rm km~s^{-1}$).\\
\end{sidewaystable*}

\end{document}